\documentclass[12pt]{article}
\usepackage{graphicx} 
\newcommand{\ra}{\rightarrow}

\newcommand{\be}{\begin{equation}}
\newcommand{\ee}{\end{equation}}
\newcommand{\bea}{\begin{eqnarray}}
\newcommand{\eea}{\end{eqnarray}}
\newcommand{\bs}{\bigskip}

\newcommand{\n}{\noindent}
\newcommand{\bc}{\begin{center}}
\newcommand{\ec}{\end{center}}
\newcommand{\bu}{\begin{underline}}
\newcommand{\eu}{\end{underline}}
\begin{document}

\bc{\Large {\bf Gauge-invariant fields in the temporal gauge, Coulomb-gauge fields,
and the Gribov ambiguity}}\ec \bs 
\bc{ Kurt Haller\\
Department of Physics, University of Connecticut\\ Storrs, Connecticut
06269}\ec \bs
\begin{abstract}
\n
We examine the relation between Coulomb-gauge fields and the gauge-invariant fields constructed in 
the temporal gauge for
two-color QCD by comparing a variety of properties, including their equal-time commutation rules
and those of their conjugate chromoelectric fields. 
We also express the temporal-gauge Hamiltonian 
in terms of gauge-invariant fields and show that it can be interpreted as a 
sum of the Coulomb-gauge Hamiltonian 
and another part that is important for determining the equations of motion of  
temporal-gauge fields, but that can never affect the time evolution of 
``physical'' state vectors.
We also discuss multiplicities of gauge-invariant temporal-gauge fields that belong to 
different topological sectors and that, in previous work, were shown to be based on the same 
underlying gauge-dependent temporal-gauge fields. We argue that these multiplicities of 
gauge-invariant fields are manifestations of the Gribov ambiguity. We show that the 
differential equation that bases the multiplicities of gauge-invariant fields 
on their underlying gauge-dependent temporal-gauge fields has nonlinearities identical to those of the 
``Gribov'' equation, which
demonstrates the non-uniqueness of Coulomb-gauge fields. These 
multiplicities of gauge-invariant fields --- and, hence, Gribov copies --- 
appear in the temporal gauge, but only with the imposition of Gauss's law and the 
implementation of gauge invariance; they do
 not arise when the theory is represented in terms of gauge-dependent fields and 
Gauss's law is left unimplemented.
\end{abstract}
\section{Introduction}
\label{sec:intro}
In earlier work, we have implemented the non-Abelian Gauss's law that applies in QCD by constructing
states that are annihilated by the ``Gauss's law operator'' ${\hat {\cal G}}^{a}({\bf r})$
for the temporal ($A^c_0=0$) gauge,~\cite{CBH2} where, for two-color QCD,
\begin{equation}
{{\cal G}}^{a}({\bf r})=
\partial_j\Pi_{j}^{a}({\bf r})+g\epsilon^{abc}A_{j}^{b}({\bf r})
\Pi_{j}^{c}({\bf r}),\;\;{\hat {\cal G}}^{a}({\bf r})={{\cal G}}^{a}({\bf r})+j_{0}^{a}({\bf r}),\;\;
\mbox{and}\;\;j^a_{0}({\bf{r}})=
g\,\,\psi^\dagger({\bf{r}})\,
{\textstyle\frac{\tau^a}{2}}\,\psi({\bf{r}})
\label{eq:quark}
\end{equation}
and where $\Pi_{j}^{a}({\bf r})$ is the negative chromoelectric field  as well as the 
momentum conjugate to the gauge field  $A_{j}^{a}({\bf r})$.
We have, furthermore, used the gauge-invariant quark and gluon  operator-valued  fields 
constructed in Ref.~\cite{CBH2} to 
transform the QCD Hamiltonian into a form in which it is expressed 
in terms of these gauge-invariant fields.~\cite{BCH3,CHQC} 
Most recently, we have studied the relation of the gauge-invariant to the gauge-dependent gauge fields  
in the temporal gauge. In particular, we have solved the nonlinear integral equation that expresses 
the requirement that the non-Abelian Gauss's law be implemented, 
and have discussed the consequences that the solutions of this integral equation have 
for the topology of the gauge-invariant gauge field.\cite{HCC} \bs

In this paper we will address a number of questions that pertain to the relation between QCD in the 
Coulomb gauge and our formulation in which QCD in the temporal (Weyl) 
gauge is expressed entirely in terms of gauge-invariant 
fields --- $i.\, e.$ in terms of operator-valued fields that commute with 
the generator displayed in Eq. (\ref{eq:quark}).
QCD in the Coulomb gauge has been discussed by Schwinger,~\cite{schwingera,schwingerb} by Christ and Lee,
~\cite{christlee,lee} by Creutz $et.\,al.,$~\cite{creutz} and by Sakita and 
Gervais,~\cite{sakita} among others. Quantization of 
QCD in the Coulomb gauge encounters a number of difficulties: These include 
the well-known Gribov ambiguity that becomes an impediment 
when Gauss's law is inverted to solve for the timelike component of the gauge field.~\cite{gribov,gribovb} 
Attention has also been called to operator ordering problems encountered when noncommuting fields 
appear multiplicatively in the Coulomb-gauge QCD Hamiltonian.~\cite{christlee,sakita}  
Some authors have circumvented the latter problem by treating the $A_0=0$ gauge fields 
as a set of Cartesian coordinates and the Coulomb-gauge fields as a set 
of curvilinear coordinates and using standard 
methods to transform from the former to the latter.~\cite{christlee,creutz}\bs

We will show in this work that the gauge-invariant fields we have constructed --- 
the gauge fields as well as the 
chromoelectric field --- have properties that so closely match those of the 
corresponding Coulomb-gauge fields,
that they can be identified with them. The gauge-invariant gauge field we 
have constructed is transverse, as 
is the gauge field in the Coulomb gauge. The gauge-invariant fields
obey commutation relations that are the same as those in the 
Coulomb gauge with the exception of operator order in the commutator.
We will, in this work, construct a number of operators that are important 
in QCD dynamics --- the Hamiltonian, 
the Gauss's law operator, the gluon and quark color-charge densities, etc. --- 
so that they are expressed entirely in terms
of these gauge-invariant fields (for short, we will call them the ``gauge-invariant'' operators). 
We will show that this gauge-invariant Hamiltonian is not the same 
as the Coulomb-gauge Hamiltonian, but that it contains a version of the 
latter accompanied by another term that is 
specific to the temporal gauge. This additional term in the gauge-invariant 
temporal-gauge Hamiltonian has no dynamical
consequences in the subspace to which the physical state variables are restricted, so 
that the two operators --- the Coulomb-gauge
Hamiltonian and the gauge-invariant Hamiltonian --- 
have identical physical consequences.  However, the additional 
term in the gauge-invariant temporal-gauge Hamiltonian affects the equations of motion
of the gauge and quark fields. Finally, we will also discuss the relation 
between the multiple solutions of the equations for the 
gauge-invariant fields that we obtained in Ref.\cite{HCC} and the 
Gribov copies of the Coulomb-gauge fields.\bs

The plan for this paper is as follows: In Section \ref{sec:gaugeinv},
we will compare the commutation rules for the Coulomb-gauge field  with those for
the gauge-invariant field that we constructed in the temporal gauge, and discuss the sense in which these two
fields can be identified with each other. We will also make a similar comparison for 
the gauge-invariant momentum (and negative chromoelectric field). 
In this section, we will also construct a number of gauge-invariant 
operator-valued quantities, such as the gauge-invariant gluon and quark 
color-charge and color-current densities, and the 
gauge-invariant version of the Gauss's law operator given in Eq. (\ref{eq:quark}). 
We will, furthermore, express the temporal-gauge Hamiltonian in terms of these gauge-invariant variables, 
and discuss its relation to the Coulomb-gauge Hamiltonian. In Section \ref{sec:gribov}, we will discuss the
multiple solutions of the nonlinear equation that determines the gauge-invariant field,  and their
relation to the Gribov copies of the Coulomb-gauge field. In 
Section \ref{sec:disc}, we will conclude with observations
based on results presented in earlier sections.\bs

\section{Gauge-invariant temporal-gauge fields}
\label{sec:gaugeinv}
The quantization of QCD in the temporal gauge avoids many of the problems encountered in quantizing
QCD in the Coulomb gauge, 
but at the expense of leaving Gauss's law still to 
be implemented after the quantization has been carried out. 
In quantizing QCD, we make use of the Lagrangian for two-color QCD, in which the gauge fields are in the 
adjoint representation of SU(2):\footnote{we 
use nonrelativistic notation, in which all space-time indices are subscripted and designate contravariant 
components of contravariant quantities 
such as $A_i^a$ or $j_i^a$, and covariant components of covariant quantities such as 
$\partial_i$. Repeated indices are summed from $1{\rightarrow}3$.}
\begin{equation}
{\cal L}=-\frac{1}{4}F_{ij}^aF_{ij}^a+\frac{1}{2}F_{i0}^aF_{i0}^a+
j_i^aA_i^a-j_0^aA_0^a+{\cal L}_{\mbox{gauge}}-{\bar \psi}
\left(m-i{\gamma}\cdot\partial\right)\psi
\label{eq:lag}
\end{equation}
where 
\begin{equation}
F_{ij}^a=\partial_jA_i^a-\partial_iA_j^a-g{\epsilon}^{abc}A_i^bA_j^c\,,\;\;\,
F_{i0}^a=\partial_0A_i^a+\partial_iA_0^a+g{\epsilon}^{abc}A_i^bA_0^c\,,
\label{eq:F}
\end{equation} 
\begin{equation}
\mbox{with}\;\;\;\;j_i^a=g\psi^{\dagger}{\alpha}_i\frac{{\tau}^a}{2}\psi,\;\;\;\mbox{and}
\;\;\;j_0^a=g\psi^{\dagger}\frac{{\tau}^a}{2}\psi\,;
\label{eq:j}
\end{equation}
${\cal L}_{\mbox{gauge}}$ is a gauge-fixing term.
When the gauge-fixing term $-A_0^aG^a$ 
is used in Eq. (\ref{eq:lag}), and the Dirac-Bergmann method of constrained 
quantization is used,\cite{dirac,bergmann} 
the Lagrange multiplier field $G^a$ is incorporated 
into the time-derivative of ${\Pi}_0^a$, which, in this case is  
$D_i\Pi_i^a+j_0^a+G^a$. The presence of the Lagrange multiplier field in the secondary constraint 
$D_i\Pi_i^a+j_0^a+G^a=0$ terminates the 
chain of secondary constraints very quickly, and 
leads to Dirac commutators that differ in only trivial ways 
from canonical Poisson commutators. However, the Gauss's law constraint is not imposed in that process.
Alternatively, it is possible to entirely avoid the need to consider primary constraints in the 
temporal gauge  by using the gauge-fixing term $-{\partial}_0A_0^aG^a$ 
instead of $-A_0^aG^a$, so that $-G^a$ becomes the 
momentum canonically conjugate to $A_0$. The gauge constraint then is ${\partial}_0A_0=0$, which, with the 
imposition of $A_0=0$ and Gauss's law at one particular time, implements both, the gauge condition and 
Gauss's law for all times;\cite{khymtemp} the same procedure can be extended to all axial gauges
for which the gauge condition is $A_0^a+{\gamma}A_3^a=0$, where $\gamma$ is a variable real 
parameter.\cite{khgenax} Even in these cases, however, the implementation of Gauss's law 
at one time --- at $t=0$, for example --- is still necessary. Finally, a very direct way of quantizing 
QCD in the temporal gauge is simply to set $A_0^a=0$ in the original Lagrangian.
 Gauss's law is not one of the Euler-Lagrange equations in this formulation, and must be imposed 
after the basic quantization has been carried out.\cite{goldjack,jackiw,rossinp} \bs

Ref. \cite{CBH2} addresses the imposition of Gauss's law in the temporal gauge by 
explicitly constructing the states that are annihilated by the non-Abelian Gauss's law operator 
given in Eq. (\ref{eq:quark}). The mathematical apparatus required for that purpose also 
enables us to construct the gauge-invariant gauge and quark fields.
This apparatus includes the defining equation for 
a nonlocal operator-valued functional of the gauge field --- the so-called ``resolvent field'' 
$\overline{{\cal{A}}_{i}^{\gamma}}({\bf{r}})$ --- which has  a central role in this 
construction. In the two-color SU(2) version of QCD, with which we are concerned in this work, 
the resolvent field appears in the gauge-invariant gluon field in the form
\be
[\,A_{{\sf GI}\,i}^{b}({\bf{r}})\,{\textstyle\frac{\tau^b}{2}}\,]
=V_{\cal{C}}({\bf{r}})\,[\,A_{i}^b({\bf{r}})\,
{\textstyle\frac{\tau^b}{2}}\,]\,
V_{\cal{C}}^{-1}({\bf{r}})
+{\textstyle\frac{i}{g}}\,V_{\cal{C}}({\bf{r}})\,
\partial_{i}V_{\cal{C}}^{-1}({\bf{r}}),
\label{eq:AdressedAxz}
\end{equation}
where $V_{\cal{C}}({\bf{r}})$ incorporates the resolvent field, as shown by 
\be
V_{\cal{C}}({\bf{r}})=
\exp\left(\,-ig{\overline{{\cal{Y}}^\alpha}}({\bf{r}})
{\textstyle\frac{\tau^\alpha}{2}}\,\right)\,
\exp\left(-ig{\cal X}^\alpha({\bf{r}})
{\textstyle{\frac{\tau^\alpha}{2}}}\right)
\label{eq:el1}
\ee
\begin{equation}
\mbox{and}\;\;\;\;V_{\cal{C}}^{-1}({\bf{r}})=
\exp\left(ig{\cal X}^\alpha({\bf{r}})
{\textstyle\frac{\tau^\alpha}{2}}\right)\,
\exp\left(\,ig{\overline{{\cal{Y}}^\alpha}}({\bf{r}})
{\textstyle\frac{\tau^\alpha}{2}}\,\right)
\label{eq:eldagq1}
\end{equation} 
\begin{equation}
\mbox{with}\;\;\;\;{\cal{X}}^\alpha({\bf{r}}) =
{\textstyle\frac{\partial_i}{\partial^2}}A_i^\alpha({\bf{r}})\;\;\;\;\mbox{and}\;\;\;\;
\overline{{\cal Y}^{\alpha}}({\bf r})=
{\textstyle \frac{\partial_{j}}{\partial^{2}}
\overline{{\cal A}_{j}^{\alpha}}({\bf r})}.
\label{eq:defXY}
\end{equation}
The composition law for two successive rotations can be used to express 
$V_{\cal{C}}({\bf{r}})$ in the form
\be
V_{\cal{C}}({\bf{r}})=
\exp\left(-ig{\cal Z}^\alpha({\bf{r}})
{\textstyle\frac{\tau^\alpha}{2}}\right)
\label{eq:vcz}
\ee
where ${\cal Z}^\alpha({\bf{r}})$ is a well-known functional of ${\cal{X}}^\alpha({\bf{r}})$ and 
$\overline{{\cal Y}^{\alpha}}({\bf r})$. 
The transversality of the gauge-invariant field 
is manifested most directly by transforming Eq. (\ref{eq:AdressedAxz}) into
\begin{equation}
A_{{\sf GI}\,i}^b({\bf{r}})=
A_{T\,i}^b({\bf{r}}) +
[\delta_{ij}-{\textstyle\frac{\partial_{i}\partial_j}
{\partial^2}}]\overline{{\cal{A}}^b_{j}}({\bf{r}}),
\label{eq:agidefin}
\ee
demonstrating that $A_{{\sf GI}\,i}^b({\bf{r}})$ is the sum of  $A_{T\,i}^b({\bf{r}})$ --- 
the transverse part of the gauge field $A_i^b({\bf{r}})$ in the temporal gauge ---
and the transverse part of the resolvent field. The resolvent field is also required for defining  the
 gauge-invariant quark field 
\begin{equation}
{\psi}_{\sf GI}({\bf{r}})=V_{\cal{C}}({\bf{r}})\,\psi ({\bf{r}})
\;\;\;\mbox{\small and}\;\;\;
{\psi}_{\sf GI}^\dagger({\bf{r}})=
\psi^\dagger({\bf{r}})\,V_{\cal{C}}^{-1}({\bf{r}})\;.
\label{eq:psiqcdg1}
\end{equation}
$V_{\cal{C}}({\bf{r}})$ can be understood as an extension, to non-Abelian gauge theories, 
of a much simpler but similar
operator that made charged QED states gauge-invariant and that was introduced by Dirac.\cite{diracgauge} 
Dirac-like operators implement gauge invariance without introducing path dependence, and it
has been argued that such Dirac-like operators have advantages for implementing 
gauge invariance over other procedures
that generate gauge-invariant charged fields with path-dependent line integrals.\cite{LMPR,haag} 
We have noted that Eqs. (\ref{eq:AdressedAxz}) and (\ref{eq:psiqcdg1}) 
implement gauge invariance --- they do not describe gauge 
transformations.\cite{CBH2} By construction, $V_{\cal{C}}({\bf{r}})$ 
transforms $A^a_i$ and $\psi$ so that they become invariant to further time-independent 
gauge-transformations consistent with the temporal gauge condition.\footnote{Gauge transformations 
within the temporal gauge, 
for which the Gauss's law operator given in Eq. (\ref{eq:quark}) is the generator, are restricted 
to time-independent gauge functions, so that the temporal-gauge constraint is not violated. }  
Although formally 
Eqs. (\ref{eq:AdressedAxz}) and (\ref{eq:psiqcdg1}) are very suggestive of gauge 
transformations of gauge and spinor fields respectively by $c$-number functions, 
that is not what they are. $V_{\cal{C}}({\bf{r}})$
is itself a functional of gauge fields, and is transformed by the same gauge transformations that
transform the fields on which it acts, just as is the case in the 
corresponding operator Dirac proposed for QED.~\cite{diracgauge}
We will show, in this section, that Eq. (\ref{eq:AdressedAxz}) describes a gauge field
that can be identified with the Coulomb-gauge field; that, in 
combination with other suitably constructed gauge-invariant quantities, it can be used in the 
representation of the temporal-gauge Hamiltonian; and that the dynamically 
effective part of this Hamiltonian can be interpreted as 
the Coulomb-gauge Hamiltonian.\bs

Eqs. (\ref{eq:AdressedAxz})-(\ref{eq:psiqcdg1}) show that the resolvent field 
$\overline{{\cal{A}}^b_{j}}({\bf{r}})$ has a central role in the representation of QCD 
in terms of gauge-invariant field variables. $\overline{{\cal{A}}^b_{j}}({\bf{r}})$ is determined by
a nonlinear integral equation that was obtained in the course of constructing the states that implement
the non-Abelian Gauss's law.\cite{CBH2} In subsequent work, this nonlinear integral equation was 
transformed (subject to an {\em ansatz}) into a nonlinear differential equation
that we solved, resulting in nonperturbative 
representations of the resolvent field and the gauge-invariant gauge field.\cite{HCC} We will discuss 
these nonperturbative solutions in more detail in section \ref{sec:gribov}.
An important corollary of the formalism that leads to this equation for the 
resolvent field is the commutation rule~\cite{chen}
  \bea
{\int}d{\bf r}\left[{\Pi}_j^b({\bf y})\,,\overline{{\cal{A}}^c_{k}}({\bf{r}})\right]&&\!\!\!\!\!\!\!\!\!
{\cal U}_{ki}^{ca}({\bf r}\!-\!{\bf x})+{\cal U}^{ba}_{ji}({\bf{y}}\!-\!{\bf x})=\nonumber \\
&&\!\!\!\!\!\!\!\!\!\!\!\!\!\!\!\!\!\!\frac{1}{2}\sum_{r=0}{\sf Tr}\left[{\tau}^bV_{\cal{C}}^{-1}({\bf{y}})
{\tau}^sV_{\cal{C}}({\bf{y}})\right]
(-1)^{r+1}g^r{\epsilon}^{\vec{v}sc}_{r}
{\textstyle\frac{\partial_j}
{{\partial}^2}}\left({\cal T}_{(r)\,k}^{\vec{v}}({\bf y}){\cal U}^{ca}_{ki}({\bf{y}}\!-\!{\bf x})\right)
\label{eq:agicomm}
\eea
\be
\mbox{where}\;\;\;{\cal U}^{ca}_{ki}({\bf{y}}-{\bf x})={\delta}_{ca}{\cal U}_{ki}({\bf{y}}-{\bf x})
\label{eq:Utrans}
\ee
\be
\mbox{with}\;\;\;\;{\cal U}_{ki}({\bf{y}}-{\bf{x}})=-i\left(\delta_{ik}-\frac{\partial_i\partial_k}
{\partial^2}\right){\delta}({\bf{y}}-{\bf{x}}).
\label{eq:uki}
\ee
In Eq.~(\ref{eq:agicomm}), we have used a notation introduced in earlier work:~\cite{CBH2,BCH3,CHQC} 
${\epsilon}^{\vec{v}ba}_{r}$ represents the chain of structure constants
 \be
{\epsilon}^{\vec{v}ba}_{r}={\epsilon}^{v_1bs_1}\,\,{\epsilon}^{s_1v_2s_2}\,{\epsilon}^{s_2v_3s_3}\,\cdots\,
\,{\epsilon}^{s_{(r-2)}v_{(r-1)}s_{(r-1)}}{\epsilon}^{s_{(r-1)}v_ra}
\label{eq:fproductN}
\ee
where repeated superscripted indices are summed from $1{\rightarrow}3$.
For $r =1$,
the chain reduces to
${\epsilon}^{\vec{v}ba}_{r}={\epsilon}^{{v}ba}$;
and for $r =0$,
${\epsilon}^{\vec{v}ba}_{r}=-\delta_{ba}$. Furthermore, we have used 
\be
\!\!\!\!{\textstyle\frac{\partial_j}
{{\partial}^2}}\left({\cal T}_{(r)\,k}^{\vec{v}}({\bf x}){\cal U}_{ki}({\bf{x}})\right)=
\left({\textstyle\frac{\partial_j}
{{\partial}^2}}
A_{{\sf GI}\;l_1}^{v_1}({\bf x})\,
{\textstyle\frac{\partial_{l_1}}{\partial^{2}}}
\left(A_{{\sf GI}\;l_2}^{v_2}({\bf x})\,
{\textstyle\frac{\partial_{l_2}}{\partial^{2}}}
\left(\cdots\left(A_{{\sf GI}\;l_{(r-1)}}^{v_{(r-1)}}({\bf x})\,
{\textstyle\frac{\partial_{l_{(r-1)}}}{\partial^{2}}}
\left((A_{{\sf GI}\;k}^{v_r}({\bf x}){\cal U}_{ki}({\bf{x}}))\right) \right)\right)\right)\right)
\label{eq:calT}
\ee
which, for $r=1$, reduces to ${\textstyle\frac{\partial_j}
{{\partial}^2}}\left({\cal T}_{(1)\,k}^{\vec{v}}({\bf x}){\cal U}_{ki}({\bf{x}})\right)=
{\textstyle\frac{\partial_j}
{{\partial}^2}}\left(A_{{\sf GI}\;k}^{v}({\bf x}){\cal U}_{ki}({\bf{x}})\,\right)$ and for 
$r=0$, reduces to ${\textstyle\frac{\partial_j}
{{\partial}^2}}\left({\cal T}_{(0)\,k}^{\vec{v}}({\bf x})
{\cal U}_{ki}({\bf{x}})\right)={\cal U}_{ji}({\bf{x})}$.
For clarity, we give the explicit form of the $r$-th element of ${\textstyle\frac{\partial_j}
{{\partial}^2}}\left({\cal T}_{(r)\,k}^{\vec{v}}({\bf y}){\cal U}_{ki}({\bf{y}}-{\bf{x}})\right)$, which is:
\bea
&&\!\!\!\!\!\!\!\!\!\!\!\!\!\!\!\!{\textstyle\frac{\partial_j}
{{\partial}^2}}\left({\cal T}_{(r)\,k}^{\vec{v}}({\bf y})
{\cal U}_{ki}({\bf{y}}-{\bf{x}})\right)\!=i(-1)^{r}
\frac{\partial}{{\partial}y_j}{\int}\frac{d{\bf z}({\scriptstyle 1})}{4{\pi}|{\bf y}-
{\bf z}({\scriptstyle 1})|}A_{{\sf GI}\,l_1}^{v_1}
({\bf z}({\scriptstyle 1}))\frac{\partial}{{\partial}z_{{l_1}}({\scriptstyle 1})}{\int}
\frac{d{\bf z}({\scriptstyle 2})}{4{\pi}|{\bf z}({\scriptstyle 1})-{\bf z}({\scriptstyle 2})|}
\times\nonumber \\
&&A_{{\sf GI}\,l_2}^{v_2}
({\bf z}({\scriptstyle 2}))\frac{\partial}{{\partial}z_{l_2}({\scriptstyle 2})}\;\;{\cdots}\;\,
{\int}\frac{d{\bf z}({\scriptstyle r-1})}{4{\pi}|{\bf z}({\scriptstyle r-2})-
{\bf z}({\scriptstyle r-1})|}A_{{\sf GI}\,{l_{r-1}}}^{v_{r-1}}({\bf z}({\scriptstyle r-1}))
\frac{\partial}{{\partial}z_{l_{r-1}}{(\scriptstyle r-1})}\times\nonumber\\
&&\left(\frac{1}{4{\pi}|{\bf z}({\scriptstyle r-1})-{\bf x}|}A_{{\sf GI}\,_i}^{v_{r}}({\bf x})
+{\int}d{\bf z}\frac{1}{4{\pi}|{\bf z}({\scriptstyle r-1})-{\bf z}|}A_{{\sf GI}\,k}^{v_r}({\bf z})
\frac{\partial}{{\partial}z_k}\frac{\partial}{{\partial}z_i}
\frac{1}{4{\pi}|{\bf z}-{\bf x}|}\right);
\label{eq:JA}
\eea
the leading ($0$-th) order term is 
\bea
{\textstyle\frac{\partial_j}
{{\partial}^2}}\left({\cal T}_{(0)\,k}^{\vec{v}}({\bf y}){\cal U}_{ki}({\bf{y}}-{\bf{x}})\right)=
-i\left(\delta_{ij}-\frac{\partial_i\partial_j}{\partial^2}\right){\delta}({\bf{y}}-{\bf{x}})\,.
\label{eq:JAzero}
\eea
We  observe that ${\cal U}^{ca}_{ki}({\bf{y}}-{\bf x})$ serves as a projection operator that 
selects the transverse parts of the fields over which it is integrated, 
and that Eqs. (\ref{eq:agidefin}) and (\ref{eq:agicomm}) in combination show that
\be
{\int}d{\bf r}\left[{\Pi}_j^b({\bf y})\,,\overline{{\cal{A}}^c_{k}}({\bf{r}})\right]
{\cal U}_{ki}^{ca}({\bf r}\!-\!{\bf x})+{\cal U}^{ba}_{ji}({\bf{y}}\!-\!{\bf x})=
\left[{\Pi}_j^b({\bf y})\,,A^a_{{\sf GI}\,i}({\bf{x}})\right].
\label{eq:agi}
\ee
The trace 
${\sf Tr}[{\tau}^bV_{\cal{C}}^{-1}{\tau}^sV_{\cal{C}}]$, which appears on the right-hand-side 
of Eq. (\ref{eq:agicomm}),  can be expressed as
$({\tau}^b)_{qn}(V_{\cal{C}}^{-1}{\tau}^sV_{\cal{C}})_{nq}$; 
with the use of the identity 
\be
\tau^b_{{\kappa}l}\tau^b_{qn}=2{\delta}_{{\kappa}n}{\delta}_{lq}-{\delta}_{{\kappa}l}{\delta}_{nq},
\label{eq:tauid}
\ee 
we obtain
\be
(\tau^b)_{{\kappa}l}({\tau}^b)_{qn}(V_{\cal{C}}^{-1}{\tau}^sV_{\cal{C}})_{nq}=
2\left(V_{\cal{C}}^{-1}{\tau}^sV_{\cal{C}}\right)_{{\kappa}l}-
{\sf Tr}\left[V_{\cal{C}}^{-1}{\tau}^sV_{\cal{C}}\right]{\delta}_{{\kappa}l}.
\label{eq:traceid}
\ee
Since ${\sf Tr}[V_{\cal{C}}^{-1}{\tau}^sV_{\cal{C}}]={\sf Tr}[{\tau}^s]=0$, it follows from 
Eqs. (\ref{eq:agicomm}) and (\ref{eq:traceid}) that
\be
\left[\left({\tau}^b\right)_{{\kappa}l}{\Pi}_j^b({\bf y})\,,A^a_{{\sf GI}\,i}({\bf{x}})\right]=
\sum_{r=0}\left(V_{\cal{C}}^{-1}({\bf{y}})
{\tau}^sV_{\cal{C}}({\bf{y}})\right)_{{\kappa}l}
(-1)^{r+1}g^r{\epsilon}^{\vec{v}sc}_{r}
{\textstyle\frac{\partial_j}
{{\partial}^2}}\left({\cal T}_{(r)\,k}^{\vec{v}}({\bf y}){\cal U}^{ca}_{ki}({\bf{y}}\!-\!{\bf x})\right)
\label{eq:pgidev}
\ee
and that, upon contraction with $[V_{\cal{C}}({\bf{y}}){\tau}^dV^{-1}_{\cal{C}}({\bf{y}})]_{l{\kappa}}$,
 we obtain
\be
\left[{\Pi}_{{\sf GI}\,j}^d({\bf y})\,,A^a_{{\sf GI}\,i}({\bf{x}})\right]=
-i\sum_{r=0}(-1)^{r+1}g^r{\epsilon}^{\vec{v}da}_{r}
{\textstyle\frac{\partial_j}
{{\partial}^2}}\left\{{\cal T}_{(r)\,k}^{\vec{v}}({\bf y})\left(\delta_{ik}-
\frac{\partial_i\partial_k}{\partial^2}\right){\delta}({\bf{y}}-{\bf{x}})\right\}
\label{eq:pgicomm}
\ee
where
\be 
{\Pi}_{{\sf GI}\,i}^d({\bf y})={\textstyle\frac{1}{2}}{\sf Tr}[V_{\cal{C}}^{-1}({\bf{y}})
\tau^dV_{\cal{C}}({\bf{y}})\tau^b]
\Pi^{b}_{i}({\bf y})
\label{eq:piginv}
\ee 
has been identified in previous work as 
the gauge-invariant momentum conjugate to the gauge field, (and 
the negative gauge-invariant chromoelectric field).\cite{CHQC} 
The following further explanatory remark for Eq. (\ref{eq:piginv}) can be given: 
We let ${\Pi}_{{\sf GI}\,i}={\Pi}_{{\sf GI}\,i}^a\textstyle{\frac{\tau^a}{2}}$, 
$A_{{\sf GI}\,i}=A^a_{{\sf GI}\,i}\textstyle{\frac{\tau^a}{2}}$, 
$A_{i}=A^a_{i}\textstyle{\frac{\tau^a}{2}}$, 
and make use of the analogy of Eq. (\ref{eq:AdressedAxz}) 
to a gauge transformation to define a corresponding $A_{\sf GI\;0}$, 
and to observe that, since in the temporal gauge
$A_0=0$, 
$$
A_{\sf GI\;0}=-\frac{i}{g}{\cal W}_0\;\;\;\mbox{where}\;\;\;{\cal W}_0=
V_{\cal{C}}\partial_0V_{\cal{C}}^{-1}\,.
$$
We recall an operator-order we have previously found necessary, 
and impose it on the obvious definition of a ``gauge-invariant momentum 
(and negative chromoelectric  field)''
\be
{\Pi}_{{\sf GI}\,i}=||\partial_iA_{\sf GI\;0}+
\partial_0A_{\sf GI\,i}-ig\left[A_{\sf GI\,i},A_{\sf GI\;0}\right]||\,,
\label{eq:pg3}
\ee
where, using a notation introduced in Ref.\cite{CBH2}, 
bracketing between double bars denotes a normal order in which all gauge 
fields and functionals of gauge fields 
appear to the left of all momenta conjugate to gauge fields, but where
 that order is imposed only after all indicated 
commutators have been evaluated (including the commutator implied by the derivatives 
$\partial_0$ and $\partial_i$). 
We observe that
\be
||\partial_0A_{\sf GI\,i}||=||\left[V_{\cal{C}}A_iV_{\cal{C}}^{-1},
{\cal W}_0\right]+V_{\cal{C}}\partial_0A_iV_{\cal{C}}^{-1}
+\frac{i}{g}\left(\partial_0V_{\cal{C}}\partial_iV_{\cal{C}}^{-1}+
V_{\cal{C}}\partial_0\partial_iV_{\cal{C}}^{-1}\right)||\,,
\label{eq:pg4}
\ee
\be
||\partial_iA_{\sf GI\;0}||=-||\frac{i}{g}\left(\partial_iV_{\cal{C}}\partial_0V_{\cal{C}}^{-1}+
V_{\cal{C}}\partial_0\partial_iV_{\cal{C}}^{-1}\right)||\,,
\label{eq:pg5}
\ee
\be
\mbox{and}\;\;\;\;-ig\left[A_{\sf GI\,i},A_{\sf GI\;0}\right]=
-\left[V_{\cal{C}}A_iV_{\cal{C}}^{-1},{\cal W}_0\right]
-\frac{i}{g}\left[\partial_0V_{\cal{C}},\partial_iV_{\cal{C}}^{-1}\right]\,.
\label{eq:pg6}
\ee
Combining Eqs. ({\ref{eq:pg4}-\ref{eq:pg6}}), we find that 
\be
{\Pi}_{{\sf GI}\,i}=||V_{\cal{C}}\partial_0A_iV_{\cal{C}}^{-1}||=||V_{\cal{C}}{\Pi}_{i}V_{\cal{C}}^{-1}||=
V_{\cal{C}}\textstyle{\frac{\tau^b}{2}}V_{\cal{C}}^{-1}{\Pi}_{i}^b\,,
\label{eq:pg7}
\ee
which agrees with Eq. (\ref{eq:piginv}). This demonstration is not 
necessary to prove that ${\Pi}^a_{{\sf GI}\,i}$
is gauge-invariant --- the transformation properties of ${\Pi}^a_{i}$ 
and of $V_{\cal{C}}\textstyle{\frac{\tau^b}{2}}V_{\cal{C}}^{-1}$ suffice 
for that --- but it nevertheless makes the identification of ${\Pi}^a_{{\sf GI}\,i}$ as the gauge-invariant 
momentum conjugate to $A^a_{{\sf GI}\,i}$ more understandable. \bs

As we will point out later in this section, 
the commutator given in Eq. (\ref{eq:pgicomm}) agrees with the corresponding commutator
for the gauge field and its conjugate momentum in the Coulomb gauge, given by Schwinger.~\cite{schwingera}
The gauge field in the Coulomb gauge and the gauge-invariant gauge field 
$A_{{\sf GI}\,i}^{b}({\bf{r}})$ constructed 
within the temporal gauge are both transverse.\footnote{The relation between 
gauge-invariance and transversality can be inverted to produce
a perturbative representation of the gauge-invariant field, as in Refs.\cite{LMPR,LMNP}.}
In Ref.\cite{schwingera} and in our work, the momentum conjugate to the gauge field --- the negative 
chromoelectric field --- has a longitudinal component, which is needed to 
implement Gauss's law. The Coulomb-gauge 
commutation rules in Refs.~\cite{christlee,lee,creutz} differ from Eq.~(\ref{eq:pgicomm}), because in these 
works, the momentum conjugate to the Coulomb-gauge field has been defined to be transverse. \bs

We next turn to a discussion of  the  ``gauge-invariant Gauss's law operator''
---  the Gauss's law operator in which gauge-invariant fields replace the original
gauge-dependent ones --- which we define as
\be
{\hat {\cal G}}^d_{\sf GI}({\bf r})=\partial_i{\Pi}_{{\sf GI}\,i}^d({\bf r})+
g{\epsilon}^{duv}A^u_{{\sf GI}\,i}({\bf{r}})
{\Pi}_{{\sf GI}\,i}^v({\bf r})+j_{0\;\sf GI}^d({\bf r})\,.
\label{eq:gaussgi}
\ee
In particular, we want to investigate whether the use of 
the gauge-invariant Gauss's law operator as the generator of gauge transformations is
consistent with the gauge invariance of $A_{{\sf GI}\,i}^{b}({\bf{r}})$. We will demonstrate this 
consistency in this section.\bs

We will first demonstrate a simple relation between  ${\hat {\cal G}}^d_{\sf GI}({\bf r})$ and 
${\hat {\cal G}}^d({\bf r})$, from which 
\be
{\int}d{\bf r}\left[{\hat {\cal G}}^b_{\sf GI}({\bf r})\,,A^a_{{\sf GI}\,i}
({\bf x})\right]\,\delta\omega^b({\bf r})=0
\label{eq:vergi}
\ee
is an immediate consequence. We observe that 
\bea
 \partial_i{\Pi}_{{\sf GI}\,i}^d({\bf r})=&&\!\!\!\!\!\!\!\!{\textstyle\frac{1}{2}}
{\sf Tr}\left\{\tau^d\partial_i
V_{\cal{C}}({\bf{r}})\tau^bV_{\cal{C}}^{-1}({\bf{r}})\right\}{\Pi}_i^b({\bf r})+
{\textstyle\frac{1}{2}}{\sf Tr}\left\{\tau^d
V_{\cal{C}}({\bf{r}})\tau^b\partial_iV_{\cal{C}}^{-1}({\bf{r}})\right\}{\Pi}_i^b({\bf r})+\nonumber\\
&&\!\!\!\!\!\!\!\!{\textstyle\frac{1}{2}}{\sf Tr}\left\{\tau^d
V_{\cal{C}}({\bf{r}})\tau^bV_{\cal{C}}^{-1}({\bf{r}})\right\}\partial_i{\Pi}_i^b({\bf r})
\label{eq:divpigi}
\eea
and that, for $\chi_i({\bf{r}})=V_{\cal{C}}({\bf{r}})\partial_iV_{\cal{C}}^{-1}({\bf{r}})$,
\be
\partial_i{\Pi}_{{\sf GI}\,i}^d=
{\textstyle\frac{1}{2}}{\sf Tr}\left\{\tau^d\left[
V_{\cal{C}}\tau^bV_{\cal{C}}^{-1}\,,\chi_i\right]\right\}{\Pi}_i^b+
{\textstyle\frac{1}{2}}{\sf Tr}\left\{\tau^d
V_{\cal{C}}\tau^bV_{\cal{C}}^{-1}\right\}\partial_i{\Pi}_i^b.
\label{eq:divpipsi}
\ee
In an Appendix, we will show that we can set 
\be
\chi_i={\textstyle\frac{i}{2}}\tau^u{\cal P}_{ui}\;\;\;\;\mbox{and}\;\;\;\;
V_{\cal{C}}\tau^bV_{\cal{C}}^{-1}=\tau^v{\cal R}_{vb},
\label{eq:demapp}
\ee
where ${\cal P}_{ui}$ and ${\cal R}_{vb}$ are functions of gauge fields only --- they are 
independent of the canonical momentum ${\Pi}_i^a$ and also contain no further SU(2) generators --- so that
\bea
{\textstyle\frac{1}{2}}{\sf Tr}\left\{\tau^d\left[
V_{\cal{C}}\tau^bV_{\cal{C}}^{-1}\,,\chi_i\right]\right\}{\Pi}_i^b&&=
{\textstyle\frac{i}{4}}{\sf Tr}\left\{\tau^d\left[\tau^v\,,\tau^u\right]\right\}{\cal R}_{vb}
{\cal P}_{ui}{\Pi}_i^b\nonumber\\
&&=\epsilon^{duv}{\cal R}_{vb}{\cal P}_{ui}{\Pi}_i^b
\label{eq:TPpi}
\eea
and 
\be
\partial_i{\Pi}_{{\sf GI}\,i}^d=\epsilon^{duv}{\cal R}_{vb}{\cal P}_{ui}
{\Pi}_i^b+{\cal R}_{db}\partial_i{\Pi}_i^b.
\label{eq:divfinal}
\ee
Similarly, using Eqs. (\ref{eq:AdressedAxz}), (\ref{eq:piginv}), and (\ref{eq:demapp}), we can express
$g{\epsilon}^{duv}A^u_{{\sf GI}\,i}{\Pi}_{{\sf GI}\,i}^v$
as
\be
g{\epsilon}^{duv}A^u_{{\sf GI}\,i}{\Pi}_{{\sf GI}\,i}^v
=g\epsilon^{duv}
\left({\cal R}_{ua}{\cal R}_{vb}A_i^a
-{\textstyle\frac{1}{g}{\cal R}_{vb}{\cal P}_{ui}}\right)
{\Pi}_i^b
\label{eq:APigi}
\ee
and find that $\epsilon^{duv}{\cal R}_{vb}{\cal P}_{ui}{\Pi}_i^b$ terms 
in Eqs. (\ref{eq:divfinal}) and (\ref{eq:APigi}) cancel, so that the gauge-invariant Gauss's 
law operator can be expressed as
\be
{\hat {\cal G}}^d_{\sf GI}({\bf r})={\cal R}_{db}\partial_i
{\Pi}_i^b+g\epsilon^{duv}{\cal R}_{ua}{\cal R}_{vb}A_i^a{\Pi}_i^b
+{\cal R}_{db}j_0^b,
\label{eq:gausstdb}
\ee 
where, in order to obtain the last term in Eq. (\ref{eq:gausstdb}), we have used\footnote{To simplify 
this argument, we make use in this discussion of a representation defined in Ref.\cite{CBH2} --- 
the so-called ${\cal C}$-representation --- in which 
$V_{\cal C}({\bf r})\psi({\bf r})$ is the gauge-invariant 
spinor and in which $j^a_{0\,{\sf GI}}=g\psi^{\dagger}V_{\cal C}^{-1}
({\tau^a/2})V_{\cal C}\psi$ is the gauge-invariant
color charge density. Elsewhere in this paper, we have used the so-called ${\cal N}$-representation, in 
which the $V_{\cal C}$ transformation has already been 
implicitly carried out for the quark field (but {\em not} the
gauge field), so that it is the quark field $\psi$ and $j^a_0=
g\psi^{\dagger}({\tau^a/2})\psi$ that are gauge invariant.} 
\be 
j_{0\;\sf GI}^d=g{\psi}^{\dagger}V_{\cal{C}}^{-1}\tau^dV_{\cal{C}}{\psi}=
{\cal R}_{db}g{\psi}^{\dagger}\tau^b{\psi}=
{\cal R}_{db}j_0^b
\label{eq:APj0}
\ee
which we justify in the Appendix. Eq. (\ref{eq:demapp}) leads to 
\be
{\cal R}_{db}={\textstyle\frac{1}{2}}{\sf Tr}[\tau^dV_{\cal{C}}\tau^bV_{\cal{C}}^{-1}],
\label{eq:R}
\ee 
so that 
\be
\epsilon^{duv}{\cal R}_{ua}{\cal R}_{vb}={\textstyle\frac{1}{4}}\epsilon^{duv}(\tau^u)_{ij}
(\tau^v)_{kp}\left(V_{\cal{C}}\tau^aV_{\cal{C}}^{-1}\right)_{ji}
\left(V_{\cal{C}}\tau^bV_{\cal{C}}^{-1}\right)_{pk}.
\label{eq:T_T}
\ee
We make use of the identity 
\be
\epsilon^{duv}(\tau^u)_{ij}(\tau^v)_{kp}=i\left(\delta_{jk}(\tau^d)_{ip}-
\delta_{ip}(\tau^d)_{kj}\right)
\label{eq:epsSU2}
\ee
to obtain 
\be
\epsilon^{duv}{\cal R}_{ua}{\cal R}_{vb}={\textstyle\frac{i}{4}}{\sf Tr}\left\{\tau^d
V_{\cal{C}}\left[\tau^b\,,\tau^a\right]V_{\cal{C}}^{-1}\right\}={\epsilon}^{baq}{\cal R}_{dq},
\label{eq:T_Tcomm}
\ee
and after relabeling dummy indices, observe that Eqs. (\ref{eq:gausstdb}) and 
(\ref{eq:T_Tcomm}) lead to 
\be
{\hat {\cal G}}^d_{\sf GI}={\cal R}_{db}\left(\partial_i{\Pi}_i^b+g\epsilon^{buv}A_i^u{\Pi}_i^v+j_0^b\right)=
{\cal R}_{db}{\hat {\cal G}}^b.
\label{eq:gaussgifinal}
\ee 
Since ${\cal R}_{ab}$ trivially commutes with $A_{{\sf GI}\,i}^b$, the fact that ${\hat {\cal G}}^d$
commutes with $A_{{\sf GI}\,i}^b$ is sufficient for the demonstration that ${\hat {\cal G}}^d_{\sf GI}$
also commutes with $A_{{\sf GI}\,i}^b$, thus proving Eq. (\ref{eq:vergi}).\bs

Another, very direct demonstration of Eq. (\ref{eq:vergi}) begins with 
the use of Eq. (\ref{eq:pgicomm}) to obtain
\be
{\int}d{\bf r}\left[\partial_j{\Pi}_{{\sf GI}\,j}^b({\bf r})\,,A^a_{{\sf GI}\,i}({\bf{x}})\right]
\delta\omega^b({\bf r})=
{\int}d{\bf r}\sum_{r=0}(-1)^{r+1}g^r{\epsilon}^{\vec{v}ba}_{r}
\left({\cal T}_{(r)\,k}^{\vec{v}}({\bf r}){\cal U}_{ki}({\bf{r}}\!-\!{\bf x})\right)
\delta\omega^b({\bf r})\,;
\label{eq:divpigicom}
\ee 
we observe that the $r=0$ term on the right-hand-side of Eq. (\ref{eq:divpigicom}) vanishes, 
because  the degenerate values listed immediately preceding and following Eq. (\ref{eq:calT}) 
demonstrate that $\partial_j({\textstyle\frac{\partial_j}
{{\partial}^2}}{\cal T}_{(0)\,k}^{\vec{v}}({\bf r}){\cal U}_{ki})=0$. We can therefore begin the sum
with $r=1$ instead of with $r=0$, redefine the dummy index $r$ to be $r+1$, and then initiate
the sum with $r=0$ for the new index $r$, obtaining 
\be
{\int}d{\bf y}\left[\partial_j{\Pi}_{{\sf GI}\,j}^b({\bf y})\,,A^a_{{\sf GI}\,i}({\bf{x}})\right]
\delta\omega^b({\bf y})=
{\int}d{\bf y}\sum_{r=0}(-1)^rg^{r+1}{\epsilon}^{\vec{v}ba}_{r+1}
\left({\cal T}_{{r+1}\,k}^{\vec{v}}({\bf y}){\cal U}_{ki}({\bf{y}}\!-\!{\bf x})\right)
\delta\omega^b({\bf y}).
\label{eq:divpiginew}
\ee
We can also evaluate 
\bea
g{\int}d{\bf y}\epsilon^{bcu}A_{{\sf GI}\,j}^c({\bf y})\left[{\Pi}_{{\sf GI}\,j}^u({\bf y})\,,
\right.&&\!\!\!\!\!\!\!\!\!\!\!\left.A^a_{{\sf GI}\,i}({\bf{x}})\right]\,\delta\omega^b({\bf y})=
{\int}d{\bf y}\sum_{r=0}(-1)^{r+1}g^{r+1}\epsilon^{bcu}{\epsilon}^{\vec{v}ua}_{r}\,{\times}\nonumber\\
&&A_{{\sf GI}\,j}^c({\bf y})\frac{\partial_j}
{{\partial}^2}\left({\cal T}_{(r)\,k}^{\vec{v}}({\bf y}){\cal U}_{ki}({\bf{y}}\!-\!{\bf x})\right)
\delta\omega^b({\bf y})
\label{eq:APicom}
\eea
and observe that 
\be
\epsilon^{bcu}{\epsilon}^{\vec{v}ua}_{r}A_{{\sf GI}\,j}^c({\bf y})\frac{\partial_j}
{{\partial}^2}\left({\cal T}_{(r)\,k}^{\vec{v}}({\bf y}){\cal U}_{ki}({\bf{y}}\!-\!{\bf x})\right)
={\epsilon}^{\vec{v}ba}_{r+1}\left({\cal T}_{(r+1)\,k}^{\vec{v}}({\bf y}){\cal U}_{ki}
({\bf{y}}\!-\!{\bf x})\right)
\label{eq:eps}
\ee
so that 
\be
\!\!\!\!\!\!\!\!g\!{\int}d{\bf y}\epsilon^{bcu}A_{{\sf GI}\,j}^c({\bf y})\left[{\Pi}_{{\sf GI}\,j}^u({\bf y})\,
A^a_{{\sf GI}\,i}({\bf{x}})\right]\,\delta\omega^b({\bf y})\!=\!\!\!{\int}d{\bf y}\sum_{r=0}(-1)^{r+1}g^{r+1}
{\epsilon}^{\vec{v}ba}_{r+1}\left({\cal T}_{(r+1)\,k}^{\vec{v}}({\bf y}){\cal U}_{ki}
({\bf{y}}\!-\!{\bf x})\right)\delta\omega^b({\bf y}).
\label{eq:APicomnew}
\ee
Cancellation between the right-hand sides of Eqs. (\ref{eq:divpiginew}) and (\ref{eq:APicomnew}) 
verifies Eq. (\ref{eq:vergi}), and therefore confirms that the use of ${\hat {\cal G}}^d_{\sf GI}$  
as the generator of infinitesimal gauge transformations is consistent with 
the gauge invariance of $A^a_{{\sf GI}\,i}({\bf{x}})$ (but not with the gauge invariance of 
$\Pi^{a}_{{\sf GI}\,i}({\bf y})\;)$.  \bs

The mathematical apparatus we developed for constructing gauge-invariant fields enables us to express the QCD 
Hamiltonian entirely in terms of these gauge-invariant fields and ${\Pi}_{{\sf GI}\,j}^b$ 
(the gauge-invariant  negative chromoelectric field).~\cite{BCH3,CHQC} The QCD Hamiltonian, represented in this way, has the 
form
\be
{\hat H}_{\sf GI}=H_{\sf GI}+H_{\cal G}
\label{eq:GHQCDN}
\ee
where $H_{\cal G}$ annihilates states that implement Gauss's law and has no dynamical
consequences in QCD. $H_{\sf GI}$ is the effective Hamiltonian in this representation of QCD and is given by
\begin{equation}
H_{\sf GI}=\int d{\bf r}\left[ \ {\textstyle \frac{1}{2}}
\Pi^{a\,{\dagger}}_{{\sf GI}\,i}({\bf r})\Pi^{a}_{{\sf GI}\,i}({\bf r})
+  {\textstyle \frac{1}{4}} F_{{\sf GI}\,ij}^{a}({\bf r}) F_{{\sf GI}\,ij}^{a}({\bf r})+
{\psi^\dagger}({\bf r})\left(\beta m-i\alpha_{i}
\partial_{i}\right)\psi({\bf r})\right] + \tilde{H}^{\prime}\,,
\label{eq:HQCDN}
\end{equation}
where, in this case, $\psi$ and ${\psi^\dagger}$ denote the gauge-invariant spinor (quark) fields, and where
\begin{equation}
F_{{\sf GI}\,ij}^{a}({\bf r})=\partial_jA_{{\sf GI}\,i}^{a}({\bf r})-\partial_iA_{{\sf GI}\,j}^{a}({\bf r})-
g\epsilon^{abc}A_{{\sf GI}\,i}^{b}({\bf r})A_{{\sf GI}\,j}^{c}({\bf r}).
\label{eq:FIJGI}
\end{equation}
$\tilde{H}^{\prime}$ is given by 
\be
\tilde{H}^{\prime}=\int d{\bf r}\left({\textstyle\frac{1}{2}}J_{0\,({\sf GI})}^{a\,\dagger}({\bf r}) 
\frac{1}{\partial^2}{\cal K}_0^a({\bf r})+
{\textstyle\frac{1}{2}}\,{\cal K}_0^a({\bf r})\frac{1}{\partial^2}J_{0\,({\sf GI})}^{a}
-{\textstyle\frac{1}{2}}{\cal K}_0^a({\bf r})\frac{1}{\partial^2}{\cal K}_0^a({\bf r})-
j^a_i({\bf r})A^a_{{\sf GI}\,i}({\bf{r}})\right)
\label{eq:Hprime}
\ee
\bs
\be
\mbox{where}\;\;\; J_{0\,({\sf GI})}^{a}=g{\epsilon}^{abc}A^b_{{\sf GI}\,i}({\bf{r}})
{\Pi}_{{\sf GI}\,i}^c({\bf r})
\label{eq:Jcolor}
\ee
is the gauge-invariant glue color-charge density, 
$j^a_i({\bf r})=g\psi^{\dagger}({\bf r}){\alpha}_i{\textstyle
\frac{\tau^a}{2}}\psi({\bf r})$ is the 
gauge-invariant quark color-current density in this representation, and where
\be
{\cal K}_0^d({\bf r})=\sum_{n=0}{\epsilon}^{\vec{\delta}d h}_{(n)}(-1)^{n}g^{n}
\left({\cal T}_{(n)}^{\vec{\delta}}({\bf r})j_0^{h}({\bf r})\right)\,,
\label{eq:Kser}
\ee
 \begin{equation}
\mbox{with}\;\;{\cal T}_{(n)}^{\vec{\delta}}({\bf r})j_0^{a}({\bf{r}})=
A_{{\sf GI}\,j(1)}^{{\delta}(1)}({\bf r})\,
{\textstyle\frac{\partial_{j(1)}}{\partial^{2}}}
\left(A_{{\sf GI}\,j(2)}^{{\delta}(2)}({\bf r})\,
{\textstyle\frac{\partial_{j(2)}}{\partial^{2}}}
\left(\cdots\left(A_{{\sf GI}\,j(n)}^{{\delta}(n)}({\bf r})\,
{\textstyle\frac{\partial_{j(n)}}{\partial^{2}}}
\left(j_0^{a}({\bf{r}})\right) \right)\right)\right).
\label{eq:calTgi}
\end{equation}
The more transparent explicit form of ${\cal K}_0^d({\bf r})$ is:
\begin{eqnarray} 
{\cal K}_0^b({\bf r})&&=
-j^b_{0}({\bf r})+g{\epsilon}^{{v}_{(1)}ba}
{A_{{\sf GI}\,i}^{{v}_{(1)}}({\bf{r}})\frac{\partial}{{\partial}r_i}
\int\frac{d{\bf x}}{4{\pi}|{\bf r}-{\bf x}|}\,j^a_{0}({{\bf x}}})+\nonumber \\
&&g^2{\epsilon}^{{v}_{(1)}bs_{(1)}}{\epsilon}^{s_{(1)}{v}_{(2)}a}{A}_{{\sf GI}\,i}^
{{v}_{(1)}}({\bf{r}})\frac{\partial}{{\partial}r_i}
\int\frac{d{\bf y}}{4{\pi}|{\bf r}-{\bf y}|}\,{A}_{{\sf GI}\,j}^{{v}_{(2)}}({\bf y})
\frac{\partial}{{\partial}y_j}\int\frac{d{\bf x}}
{4{\pi}|{\bf y}-{\bf x}|}\,j^a_{0}({\bf x})+\cdots\nonumber \\
+&&g^n{\epsilon}^{{v}_{(1)}bs_{(1)}}{\cdots}{\epsilon}^{s_{(n-2)}{v}_{(n-1)}s_{(n-1)}}
{\epsilon}^{s_{(n-1)}{v}_{(n)}a}{A}_{{\sf GI}\,i}^{{v}_{(1)}}({\bf{r}})\frac{\partial}{{\partial}r_i}
\int\frac{d{\bf y}_{(1)}}{4{\pi}|{\bf r}-{\bf y_{(1)}}|}\,{\cdots}
\times\nonumber \\ 
&&A_{{\sf GI}\,{\ell}}^
{{v}_{(n-2)}}({\bf y}_{(n-3)})\frac{\partial}{{\partial}y_{(n-3)\;\ell}}
\int\frac{d{\bf y}_{(n-2)}}{4{\pi}|{\bf y}_{(n-3)}-{\bf y}_{(n-2)}|}\,{A}_{{\sf GI}\,j}^{{v}_{(n-1)}}
({\bf y}_{(n-2)})\frac{\partial}{{\partial}y_{(n-2)\;j}} \times \nonumber \\
&&
\int\frac{d{\bf y}_{(n-1)}}{4{\pi}|{\bf y}_{(n-2)}-{\bf y}_{(n-1)}|}{A}_{{\sf GI}\,k}^{{v}_{(n)}}
({\bf y}_{(n-1)})\,\frac{\partial}{{\partial}y_{(n-1)\;k}}\int\frac{d{\bf x}}{4{\pi}|
{\bf y}_{(n-1)}-{\bf x}|}\,
j^a_{0}({{\bf x}})+\cdots\;.
\label{eq:calTexp}
\end{eqnarray}
In formulating Eqs. (\ref{eq:HQCDN})-(\ref{eq:calTexp}), we note that while it is trivial that 
$A_{{\sf GI}i}^{a}({\bf r})$ is hermitian, and while we will show that 
${\hat {\cal G}}^b_{\sf GI}({\bf r})$ is hermitian as well,  
$\Pi^{b\,{\dagger}}_{{\sf GI}\,i}({\bf r})$ and $\Pi^{b}_{{\sf GI}\,i}({\bf r})$ are not 
identical.~\footnote{a similar point is made in Ref.\cite{creutz}; note, however, that the chromoelectric field in 
Ref.\cite{creutz} is transverse, while our chromoelectric field includes its longitudinal part, as in 
Ref.\cite{schwingera}; the longitudinal part of the chromoelectric field accounts for its lack of hermiticity.} 
As can be seen from Eq. (\ref{eq:piginv}),  
in order for $\Pi^{b\,{\dagger}}_{{\sf GI}\,i}({\bf r})$ to be hermitian, 
$\Pi^{b}_{i}({\bf r})$ would have to commute with 
$R_{db}({\bf r})$ (defined in Eq. (\ref{eq:R})).
We therefore distinguish between $\Pi^{b\,{\dagger}}_{{\sf GI}\,i}({\bf r})$ and 
$\Pi^{b}_{{\sf GI}\,i}({\bf r})$ in this discussion; similarly, Eq.~(\ref{eq:pgicomm}) shows that 
$J_0^a$ and $J_0^{a\,\dagger}$ will differ. The Hamiltonian and its components, 
${\hat H}_{\sf GI}$, $H_{\sf GI}$, $H_{\cal G}$, and $\tilde{H}^{\prime}$, are all manifestly hermitian. \bs

Eq. (\ref{eq:Hprime}) is very suggestive of an interaction Hamiltonian in the Coulomb gauge, but with
${\cal K}_0^a$ appearing in the position in which one might expect to find the charge
density of the matter field $j^a_0$. ${\cal K}_0^a$ contains $j^a_0$ as a crucial component,  but
extends the latter's dynamical effect over a greater region in space than $j^a_0$ itself occupies,
through a series of ``chains'' of interactions in which each link has the form 
${\epsilon}^{{\alpha}v{\beta}}A_{{\sf GI}\,i}^v{\textstyle}{\frac{\partial_i}{\partial^2}}$. 
The effect of these chains of interactions is that if, for example, $j^a_0$ describes quark 
color-charges that are limited to a relatively small volume, ${\cal K}_0^a$ could be significant 
over a substantially larger region of space.  Further investigation of these nonlocal
interactions can be facilitated by the observation that 
\be
{\cal K}_0^a+g{\epsilon}^{avb}A_{{\sf GI}\,i}{\textstyle}{\frac{\partial_i}{\partial^2}}{\cal K}_0^b=-j^a_0.
\label{eq:Kscr}
\ee
The issue here is not that ${\cal K}_0^a$ is gauge-invariant. In fact, in the representation used 
to express $\tilde{H}^{\prime}$, both ${\cal K}_0^a$ and $j^a_0$ are gauge-invariant, as has been discussed 
in Ref. \cite{CHQC}. \bs

 A further significant feature of $\tilde{H}^{\prime}$  
 is that, besides the nonlocal interaction  
$-{\textstyle}{\frac{1}{2}}{\cal K}_0^a{\textstyle}{\frac{1}{\partial^2}}{\cal K}_0^a$,
it also includes additional interactions of the gauge-invariant glue color-charge density $J_0^a$ 
with ${\cal K}_0^a$, 
so that quark and glue color charge-densities interact with each other in the non-Abelian theory 
in which the gauge field, as well as the matter fields to which it couples, carry color-charge.
\bs

$H_{\cal G}$ --- the part of the Hamiltonian that
annihilates states that implement Gauss's law and therefore has no dynamical consequences --- 
can be given as
\be
H_{\cal G}=-{\textstyle\frac{1}{2}}\int d{\bf r}\left[{\cal G}_{\sf GI}^{a}
\frac{1}{\partial^2}{\cal K}_0^a({\bf r})+
{\cal K}_0^a({\bf r})\frac{1}{\partial^2}{\cal G}_{\sf GI}^{a}\right]\,.
\label{eq:HamGauss}
\ee
Since ${\cal G}_{\sf GI}^{a}$ is hermitian, and since any state 
$|\Psi\rangle$ for which ${\cal G}_{\sf GI}^a({\bf{x}})\,|\Psi\rangle=0$
will time-evolve so that  ${\cal G}_{\sf GI}^a({\bf{x}})\,\exp(-i{\hat H}_{\sf GI}\,t)|\Psi\rangle=0$ 
as well --- properties of ${\cal G}_{\sf GI}^{a}$ that we will prove 
later in this section --- $H_{\cal G}$ will not contribute to matrix 
elements for physical processes or have any affect on the properties of 
states in the space of ``physical'' states which must implement the non-Abelian Gauss's law.
\bs

To demonstrate the hermiticity of ${ {\cal G}}^b_{\sf GI}({\bf r})$, we use Eq. (\ref{eq:gaussgifinal})
to show that 
\be
{\Delta}_g({\bf r})={ {\cal G}}^{d\,\dagger}_{\sf GI}({\bf r})-{ {\cal G}}^d_{\sf GI}({\bf r})=
\left[{ {\cal G}}^b({\bf r})\,,R_{db}({\bf r})\right].
\ee
From the proof that $V_{\cal{C}}({\bf r})$ transforms gauge-dependent temporal-gauge quark and gauge 
fields into gauge-invariant fields as shown in 
Eqs.~(\ref{eq:AdressedAxz}) and (\ref{eq:psiqcdg1}),\cite{mario} 
we observe that
\be
\left[{ {\cal G}}^c({\bf y})\,,V_{\cal{C}}({\bf x})\right]=gV_{\cal{C}}({\bf x})
{\textstyle \frac{\tau^c}{2}}{\delta}({\bf x}-{\bf y})\;\;\;\mbox{and}\;\;\;
\left[{ {\cal G}}^c({\bf y})\,,V^{-1}_{\cal{C}}({\bf x})\right]=-g{\textstyle \frac{\tau^c}{2}}
V^{-1}_{\cal{C}}({\bf x}){\delta}({\bf x}-{\bf y})\,,
\label{eq:Gherma}
\ee
so that 
\be
{\Delta}_g({\bf r})={\textstyle \frac{1}{2}}{\sf Tr}\left\{{\tau^b}\left[{{\cal G}}^c({\bf r})\,,
V_{\cal{C}}({\bf r}){\tau^c}V^{-1}_{\cal{C}}({\bf r})\right]\right\}=0
\label{eq:Ghermb}
\ee
and ${ {\cal G}}^d_{\sf GI}({\bf r})$ is shown to be hermitian.\footnote{We interpret the commutators 
$[{ {\cal G}}^c({\bf r})\,,V_{\cal{C}}({\bf r})]$ and $[{ {\cal G}}^c({\bf r})\,,V_{\cal{C}}^{-1}({\bf r})]$
as $\lim_{{\bf r^{\prime}}{\rightarrow}{\bf r}}[{ {\cal G}}^c({\bf r^{\prime}})\,,V_{\cal{C}}({\bf r})]$
and $\lim_{{\bf r^{\prime}}{\rightarrow}{\bf r}}[{ {\cal G}}^c({\bf r^{\prime}})\,,
V_{\cal{C}}^{-1}({\bf r})]$ respectively to regularize the $0$ argument of the delta-function.}\bs

We will examine the hermiticity of  ${\Pi}_{{\sf GI}\,i}^a({\bf x})$ by first evaluating
 $[{\Pi}_{{\sf GI}\,i}^a({\bf x})\,,{\Pi}_{{\sf GI}\,i}^b({\bf y})]$ and comparing it 
with the commutator $[{\Pi}_{{\sf GI}\,i}^{a}({\bf x})\,,{\Pi}_{{\sf GI}\,i}^{b\,\dagger}({\bf y})]$.
In this way, we avoid the necessity of evaluating the singular commutator 
$[{\Pi}_{i}^a({\bf x})\,,R_{ba}({\bf x})]$. To evaluate these commutators, we first replace 
$j_0^a({\bf r})$ by $-g{\textstyle \frac{\tau^a}{2}}{\delta}({\bf r}-{\bf x})$ 
in Eqs. (7) and (19) in Ref.\cite{BCH3}, a step justified by the fact that both sets of quantities 
obey the same closed commutator algebra. In this way, we obtain
\be
V_{\cal{C}}({\bf x}){\Pi}_{j}^b({\bf y})V^{-1}_{\cal{C}}({\bf x})={\Pi}_{j}^b({\bf y})+
\sum_{n=0}g^n(-1)^nR_{db}({\bf y}){\epsilon}_n^{{\vec \delta}dh}{\textstyle \frac{\partial_j}{{\partial}^2}}
\left({\cal T}_{(n)}^{\vec{\delta}}({\bf y})g{\textstyle \frac{\tau^h}{2}}{\delta}({\bf y}-{\bf x})\right)
\label{eq:Piherma}
\ee
and therefore that 
\be
\!\!\!\!\!\!\!\!\!\left[{\Pi}_{j}^b({\bf y})\,,V_{\cal{C}}({\bf x}){\tau^a}V^{-1}_{\cal{C}}({\bf x})\right]\!=
\sum_{n=0}g^{n+1}(-1)^{n+1}R_{db}({\bf y}){\epsilon}_n^{{\vec \delta}dh}
{\textstyle \frac{\partial_j}{{\partial}^2}}
\left\{{\cal T}_{(n)}^{\vec{\delta}}({\bf y}){\delta}({\bf y}-{\bf x})
\left[{\textstyle \frac{\tau^h}{2}}\,,V_{\cal{C}}({\bf x}){\tau^a}V^{-1}_{\cal{C}}({\bf x})\right]\right\}
\label{eq:Pihermb}
\ee
from which 
\be 
\left[{\Pi}_{j}^b({\bf y})\,,R_{{\alpha}a}({\bf x})\right]\!=
i\sum_{n=0}g^{n+1}(-1)^{n+1}R_{db}({\bf y}){\epsilon}_n^{{\vec \delta}dh}{\epsilon}^{hc{\alpha}}
{\textstyle \frac{\partial_j}{{\partial}^2}}
\left\{{\cal T}_{(n)}^{\vec{\delta}}({\bf y})
{\delta}({\bf y}-{\bf x})R_{ca}({\bf x})\right\}
\label{eq:Pihermc}
\ee
follows; and, using $R_{cb}({\bf y})R_{db}({\bf y})={\delta}_{cd}$, we also obtain
\be 
\left[{\Pi}_{{\sf GI}\,j}^{\beta}({\bf y})\,,R_{{\alpha}a}({\bf x})\right]\!=
i\sum_{n=0}g^{n+1}(-1)^{n+1}{\epsilon}_n^{{\vec \delta}{\beta}h}{\textstyle {\epsilon}^{hc{\alpha}}
\frac{\partial_j}{{\partial}^2}}\left\{{\cal T}_{(n)}^{\vec{\delta}}({\bf y})
{\delta}({\bf y}-{\bf x})R_{ca}({\bf x})\right\}\,.
\label{eq:Pihermd}
\ee
Since 
\be
\left[{\Pi}_{{\sf GI}\,i}^{\alpha}({\bf x})\,,{\Pi}_{{\sf GI}\,j}^{\beta}({\bf y})\right]=
\left[{\Pi}_{{\sf GI}\,i}^{\alpha}({\bf x})\,,R_{\beta b}({\bf y})\right]{\Pi}_{j}^b({\bf y})-
\left[{\Pi}_{{\sf GI}\,j}^{\beta}({\bf y})\,,R_{\alpha a}({\bf x})\right]{\Pi}_{i}^a({\bf x})\,,
\label{eq:Piherme}
\ee
it follows that
\bea
\left[{\Pi}_{{\sf GI}\,i}^{\alpha}({\bf x})\,,{\Pi}_{{\sf GI}\,j}^{\beta}({\bf y})\right]=
i\sum_{n=0}g^{n+1}(-1)^{n+1}&&\!\!\!\!\!\!\!\!\!\!\left[{\epsilon}_n^{{\vec \delta}{\alpha}h}
{\textstyle {\epsilon}^{h{\gamma}{\beta}}
\frac{\partial_i}{{\partial}^2}}\left\{{\cal T}_{(n)}^{\vec{\delta}}({\bf x})
{\delta}({\bf x}-{\bf y})
{\Pi}_{{\sf GI}\,j}^{\gamma}({\bf y})\right\}\right.\nonumber \\
&&\!\!\!\!\!\!\!\!\!\!-\left.{\epsilon}_n^{{\vec \delta}{\beta}h}{\textstyle {\epsilon}^{h{\gamma}{\alpha}}
\frac{\partial_j}{{\partial}^2}}\left\{{\cal T}_{(n)}^{\vec{\delta}}({\bf y})
{\delta}({\bf y}-{\bf x})
{\Pi}_{{\sf GI}\,i}^{\gamma}({\bf x})\right\}\right]\,.
\label{eq:Pihermf}
\eea
The explicit form of the $n$-th order term $\frac{\partial_j}{{\partial}^2}
\left\{{\cal T}_{(n)}^{\vec{\delta}}({\bf y})
{\delta}({\bf y}-{\bf x})
{\Pi}_{{\sf GI}\,i}^{\gamma}({\bf x})\right\}$ is
\bea
\frac{\partial_j}{{\partial}^2}\left\{{\cal T}_{(n)}^{\vec{\delta}}({\bf y})\right.&&\!\!\!\!\!\!\!\!\!\left.
{\delta}({\bf y}-{\bf x})
{\Pi}_{{\sf GI}\,i}^{\gamma}({\bf x})\right\}\!=(-1)^n
\frac{\partial}{{\partial}y_j}{\int}\frac{d{\bf z}({\scriptstyle 1})}{4{\pi}|{\bf y}-
{\bf z}({\scriptstyle 1})|}A_{{\sf GI}\,l_1}^{{\delta}_1}
({\bf z}({\scriptstyle 1}))\frac{\partial}{{\partial}z({\scriptstyle 1})_{l_1}}{\times}\nonumber \\
&&\!\!\!\!\!\!\!\!\!{\int}\frac{d{\bf z}({\scriptstyle 2})}
{4{\pi}|{\bf z}({\scriptstyle 1})-{\bf z}({\scriptstyle 2})|}
A_{{\sf GI}\,l_2}^{{\delta}_2}
({\bf z}({\scriptstyle 2}))\frac{\partial}{{\partial}z({\scriptstyle 2})_{l_2}}\;
{\cdots}{\int}\frac{d{\bf z}({\scriptstyle n})}{4{\pi}|{\bf z}({\scriptstyle n-1}))-
{\bf z}({\scriptstyle n})|}{\times}\nonumber \\    
&&A_{{\sf GI}\,l_n}^{{\delta}_n}
({\bf z}({\scriptstyle n}))\frac{\partial}{{\partial}z({\scriptstyle n})_{l_n}}
\frac{1}{{4{\pi}|{\bf z}({\scriptstyle n})-{\bf x}|}}{\Pi}_{{\sf GI}\,i}^{\gamma}({\bf x})
\label{eq:Pihermg}
\eea 
and the leading ($n=0$) term of the commutator given in Eq. (\ref{eq:Piherme}) is 
$$ig{\epsilon}^{\alpha\beta\gamma}\left(\frac{\partial}{\partial x_i}\frac{1}{{4{\pi}|{\bf x}-{\bf y}|}}
{\Pi}_{{\sf GI}\,j}^{\gamma}({\bf y})+\frac{\partial}{\partial y_j}\frac{1}{{4{\pi}|{\bf x}-{\bf y}|}
}{\Pi}_{{\sf GI}\,i}^{\gamma}({\bf x})\right)\,.$$
The delta-functions that appear in Eq. (\ref{eq:Pihermf}) eliminate the integrations over the last of the 
inverse laplacian in the chain described in Eq. (\ref{eq:calTgi}). When there is no delta-function in the 
expressions on which  ${\cal T}_{(n)}^{\vec{\delta}}$ acts, the last inverse 
laplacian is also integrated over, as 
can be seen by comparing with Eqs. (\ref{eq:Kser}) and (\ref{eq:calTexp}).\bs 

We can apply the same procedure used to obtain Eq. (\ref{eq:Pihermf}) to the commutator 
of ${\Pi}_{{\sf GI}\,i}^{\alpha}({\bf x})$ and the 
hermitian adjoint of ${\Pi}_{{\sf GI}\,j}^{\beta}({\bf y})$, in which case we get 
\be
\left[{\Pi}_{{\sf GI}\,i}^{\alpha}({\bf x})\,,{\Pi}_{{\sf GI}\,j}^{\beta\,\dagger}({\bf y})\right]=
R_{{\alpha}a}({\bf x}){\Pi}_{j}^b({\bf y})\left[{\Pi}_{i}^{\alpha}({\bf x})\,,R_{\beta b}({\bf y})\right]-
\left[{\Pi}_{{\sf GI}\,j}^{\beta}({\bf y})\,,R_{\alpha a}({\bf x})\right]{\Pi}_{i}^a({\bf x})\,,
\label{eq:Piada}
\ee
and therefore that
\bea
\left[{\Pi}_{{\sf GI}\,i}^{\alpha}({\bf x})\,,{\Pi}_{{\sf GI}\,j}^{\beta\,\dagger}({\bf y})\right]=
i\sum_{n=0}g^{n+1}(-1)^{n+1}&&\!\!\!\!\!\!\!\!\left[{\epsilon}_n^{{\vec \delta}
{\alpha}h}{\textstyle {\epsilon}^{h{\gamma}{\beta}}
{\Pi}_{{\sf GI}\,j}^{\gamma\,\dagger}({\bf y})\frac{\partial_i}
{{\partial}^2}}\left\{{\cal T}_{(n)}^{\vec{\delta}}({\bf x})
{\delta}({\bf x}-{\bf y})
\right\}\right.\nonumber \\
-&&\!\!\!\!\!\!\!\!\left.{\epsilon}_n^{{\vec \delta}{\beta}h}{\textstyle {\epsilon}^{h{\gamma}{\alpha}}
\frac{\partial_j}{{\partial}^2}}\left\{{\cal T}_{(n)}^{\vec{\delta}}({\bf y})
{\delta}({\bf y}-{\bf x})
{\Pi}_{{\sf GI}\,i}^{\gamma}({\bf x})\right\}\right]\nonumber \\
-&&\!\!\!\!\!\!\!\!\left[R_{{\beta}b}({\bf y})\,,{\Pi}_{i}^a({\bf x})\right]\,
\left[R_{{\alpha}a}({\bf x})\,,{\Pi}_{j}^b({\bf y})\right]\,.
\label{eq:Piadb}
\eea
An alternate expression for Eq. (\ref{eq:Pihermf}) can be obtained by defining
${\cal D}({\bf x},{\bf y})$ as an inverse of the Faddeev-Popov operator\footnote{In this, as well as in 
a number of similar cases, the gauge-invariant field $A_{{\sf GI}\,i}^c$ replaces the standard gauge-dependent
temporal-gauge field $A_{i}^c$ in the gauge-covariant derivative.}
$(D{\cdot}\partial)^{ab}=(\delta_{ab}+g{\epsilon}^{aub}
A_{{\sf GI}\,n}^u\textstyle{\frac{{\partial_n}}{{\partial}^2}){\partial^2}}$, given by 
\be
D{\cdot}{\partial}({\bf x}){\cal D}({\bf x},{\bf y})={\delta}({\bf x}-{\bf y}).
\label{eq:FPi}
\ee
${\cal D}({\bf x},{\bf y})$ can be expanded as a series in the form
\be
{\cal D}^{dh}({\bf x},{\bf y})=-\frac{1}{\partial^2}\sum_{n=0}{\epsilon}^{\vec{\delta}d h}_{(n)}(-1)^{n}g^{n}
{\cal T}_{(n)}^{\vec{\delta}}({\bf x}){\delta}({\bf x}-{\bf y})
\label{eq:FPser}
\ee
so that Eq. (\ref{eq:Pihermf}) can be expressed as
\be
\left[{\Pi}_{{\sf GI}\,i}^{\alpha}({\bf x})\,,{\Pi}_{{\sf GI}\,j}^{\beta}({\bf y})\right]=
ig\left\{{\partial_i}{\cal D}^{{\alpha}h}({\bf x},{\bf y}){\epsilon}^{h\gamma\beta}
{\Pi}_{{\sf GI}\,j}^{\gamma}({\bf y})-
{\partial_j}{\cal D}^{{\beta}h}({\bf y},{\bf x}){\epsilon}^{h\gamma\alpha}
{\Pi}_{{\sf GI}\,i}^{\gamma}({\bf x})\right\}.
\label{eq:compsch}
\ee
In the form given in Eq. (\ref{eq:compsch}), the equal-time commutation 
relation in Eq. (\ref{eq:Pihermf}) can be seen to be in agreement with 
the one given by Schwinger in Ref.\cite{schwingera} {\em modulo} the fact 
that Schwinger's Coulomb-gauge chromoelectric field 
operators are defined to be hermitian, whereas ours are not. 
The operator-ordering in the commutators in Ref.~\cite{schwingera} 
that correspond to our Eq.~(\ref{eq:Pihermf}) therefore differs from ours. We have defined the gauge-invariant 
chromoelectric fields so that the Hamiltonian in Eq. (\ref{eq:HQCDN}) has a simple and tractable form. 
Furthermore, with $D_i^{ba}={\delta}_{ba}\partial_i+g{\epsilon}^{bua}A_{{\sf GI}\,i}^u$, Eq. (\ref{eq:FPser}), and the identity
\bea
\left({\delta}_{ij}{\delta}_{ba}+\frac{\partial_j}{\partial^2}\sum_{n=0}
{\epsilon}^{\vec{\delta}b h}_{(n)}\right.&&\!\!\!\!\!\!\!\!\!\!\left.(-1)^{n}g^{n}
{\cal T}_{(n)}^{\vec{\delta}}({\bf x})D_i^{ha}({\bf x})\right)
{\delta}({\bf x}-{\bf y})=\nonumber\\
&&\!\!\!\!\!\!\!\!\sum_{n=0}{\epsilon}^{\vec{\delta}b a}_{(n)}(-1)^{n+1}g^{n}
\frac{\partial_j}{\partial^2}{\cal T}_{(n)\,k}^{\vec{\delta}}({\bf x})
\left(\delta_{ik}-\frac{\partial_i\partial_k}{\partial^2}\right){\delta}({\bf x}-{\bf y})
\label{eq:invid}
\eea
we can also confirm the agreement between Eq. (\ref{eq:pgicomm}) and the corresponding 
commutation rule given by Schwinger in Ref.~\cite{schwingera}, again {\em modulo} operator order 
in the corresponding expressions. Our commutators
in Eqs. (\ref{eq:pgicomm}) and (\ref{eq:Pihermf}) differ much more substantially from those 
in Refs.\cite{christlee,lee,creutz} because, in these 
references, the Coulomb-gauge chromoelectric field is defined 
to be transverse, in contrast to the practice followed
in Schwinger's treatment of the Coulomb gauge as well as in our work.
\bs

We can use Eq. (\ref{eq:Pihermc}) to substitute explicit expressions for the commutators in the last line of 
Eq. (\ref{eq:Piadb}); but comparison of Eqs. (\ref{eq:Pihermf}) and (\ref{eq:Piadb}) suffices to confirm that 
${\Pi}_{{\sf GI}\,j}^{\beta\,\dagger}({\bf y})$ differs from 
${\Pi}_{{\sf GI}\,j}^{\beta}({\bf y})$.
It is also possible
to evaluate ${\Pi}_{{\sf GI}\,j}^{\beta\,\dagger}({\bf y})-{\Pi}_{{\sf GI}\,j}^{\beta}({\bf y})$ directly. The 
resulting expression is clearly nonvanishing, but the expression is lengthy as well as singular 
and it is not necessary to quote it 
in detail in order to make the point that the gauge-invariant momentum 
(and negative chromoelectric field) is not 
hermitian.  
We observe that $\partial_i{\Pi}^a_{{\sf GI}\,i}$ and $g\epsilon^{abc}A^b_{{\sf GI}\,i}
{\Pi}^b_{{\sf GI}\,i}+j^a_{0\,{\sf GI}}$, the two component parts of ${\hat {\cal G}}^a_{\sf GI}$,
are separately gauge-invariant, but are not separately hermitian. 
\bs

It is also of interest to examine whether the commutator algebra of the gauge-invariant Gauss's law operators 
closes, by examining the commutator $[{\hat {\cal G}}^a_{\sf GI}({\bf{x}}),
\,{\hat {\cal G}}^b_{\sf GI}({\bf{y}})]$.
From Eq.~(\ref{eq:gaussgifinal}), we observe that 
\be
\!\!\!\!\left[{{\cal G}}^a_{\sf GI}({\bf{x}}),\,{\cal G}^b_{\sf GI}({\bf{y}})\right]=
{\cal R}_{ac}({\bf{x}})\left[{{\cal G}}^c({\bf{x}}),\,{\cal R}_{bd}({\bf{y}})\right]-
{\cal R}_{bd}({\bf{y}})\left[{ {\cal G}}^d({\bf{y}}),\,{\cal R}_{ac}({\bf{x}})\right]+
{\cal R}_{ac}({\bf{x}}){\cal R}_{bd}({\bf{y}})\left[{{\cal G}}^c({\bf{x}}),\,{\cal G}^d({\bf{y}})\right]
\label{eq:gbarcomm1}
\ee 
and, using 
\be
\left[{{\cal G}}^c({\bf{x}}),\,{\cal R}_{bd}({\bf{y}})\right]=
ig{\epsilon}^{cdq}{\cal R}_{bq}({\bf{y}}){\delta}({\bf{x}}-{\bf{y}}),
\label{eq:gcommr}
\ee
\be
\mbox{that}\;\;\;\left[{{\cal G}}^a_{\sf GI}({\bf{x}}),\,{\cal G}^b_{\sf GI}({\bf{y}})\right]=
-ig{\epsilon}^{abc}{{\cal G}}^c_{\sf GI}({\bf{x}}){\delta}({\bf{x}}-{\bf{y}}).
\label{eq:gbarcomm2}
\ee
This shows that the commutator algebra of the gauge-invariant Gauss's 
law operators closes and, in fact, is almost identical to the
algebra of the gauge dependent Gauss's law operators except 
for a relative sign change between the two cases. \bs 

One consequence of the time-independence of the non-Abelian 
Gauss's law operator given in Eq.~(\ref{eq:quark}) 
is the fact that $[H, {{\cal G}}^a({\bf{x}})]=0$, so that when a state implements 
Gauss's law ($i.\,e.$ when ${{\cal G}}^a({\bf{x}})\,|\Psi\rangle=0$), the time-evolved state 
${{\cal G}}^a({\bf{x}})\exp(-iHt)|\Psi\rangle$ also vanishes. It is therefore 
important to inquire whether a similar property can be
ascribed to the gauge-invariant non-Abelian Gauss's law operator ${\cal G}^a_{\sf GI}$, 
so that the set of states that implement  ${{\cal G}}^a_{\sf GI}({\bf{x}})\,|\Psi\rangle=0$ 
also implement that same constraint at all later times,  when $\exp(-i{\hat H_{\sf GI}}t)$ is the 
time evolution operator. We address this question by observing that 
\be
\left[{{\cal G}}^a_{\sf GI}({\bf{x}}),\,\Pi^b_{\sf GI}({\bf{y}})\right]
=\sum_{r=0}(-1)^{r+1}g^{r+1}{\epsilon}^{\vec{v}bh}_{r}{\epsilon}^{hqa}
{\textstyle\frac{\partial_j}
{{\partial}^2}}\left({\cal T}_{(r)}^{\vec{v}}({\bf y})
{\delta}({\bf{x}}-{\bf{y}})\right){{\cal G}}^q_{\sf GI}({\bf{x}})
\label{eq:ggitime1}
\ee
follows from Eqs. (\ref{eq:Pihermd}) and (\ref{eq:gaussgifinal}), 
and that ${{\cal G}}^a_{\sf GI}({\bf{x}})$ commutes
 with ${\cal K}^b_0({\bf{r}})$ 
and with $j^b_0({\bf{r}})$. The commutator of ${\hat H_{\sf GI}}$ 
and ${{\cal G}}^a_{\sf GI}({\bf{x}})$ therefore 
receives contributions only from commutators of ${{\Pi}}^q_{\sf GI}({\bf{y}})$ and 
${{\cal G}}^a_{\sf GI}({\bf{x}})$; and, when
$[{{\Pi}}^c_{\sf GI}({\bf{y}}),\,{{\cal G}}^a_{\sf GI}({\bf{x}})]$ 
is to the left of another operator, as in 
${\int}d{\bf{y}}[{{\Pi}}^c_{\sf GI}({\bf{y}}),\,
{{\cal G}}^a_{\sf GI}({\bf{x}})]{{\Pi}}^c_{\sf GI}({\bf{y}})$ or in 
${\int}d{\bf{y}}g{\epsilon}^{bpq}{{A}}^p_{\sf GI}({\bf{y}})[{{\Pi}}^q_{\sf GI}
({\bf{y}}),\,{{\cal G}}^a_{\sf GI}({\bf{x}})]
({\partial}^{-2}){\cal K}^b_0({\bf{y}})$, the resulting 
${{\cal G}}^a_{\sf GI}({\bf{x}})$ either commutes with the operators to its 
right, or, in moving to the right of those that do not commute with it, 
produces still other terms that have a 
${{\cal G}}^d_{\sf GI}({\bf{r}})$ on the extreme right-hand side. 
We therefore observe that the most general expression 
for the commutator of ${\hat H_{\sf GI}}$ and ${{\cal G}}^a_{\sf GI}({\bf{x}})$ is 
\be
\left[{\hat H}_{\sf GI},\,{\cal G}^a_{\sf GI}({\bf{x}})\right]=
{\int}d{\bf r}{\chi}^{ac}({\bf{r}},{\bf{x}}){\cal G}^c_{\sf GI}({\bf r})
\label{eq:ggitime2}
\ee
where ${\chi}^{ac}({\bf{r}},{\bf{x}})$ is a nonlocal functional of $\bf x$ and $\bf r$.
Although ${\cal G}^a_{\sf GI}({\bf{x}})$ is not time-independent, 
it is nevertheless true that a state $|\Psi\rangle$
for which ${\cal G}^a_{\sf GI}({\bf{x}})\,|\Psi\rangle=0$ for all 
values of $\bf x$ and $a$ will time-evolve so that  
${\cal G}^a_{\sf GI}({\bf{x}})\,\exp(-i{\hat H}_{\sf GI}\,t)|\Psi\rangle=0$ as well.\bs 

The Hamiltonian given in Eqs. (\ref{eq:GHQCDN})-(\ref{eq:HamGauss}) relates QCD in the 
temporal-gauge to its Coulomb-gauge formulation. ${\hat H}_{\sf GI}$ is the Hamiltonian of QCD in the 
temporal gauge. It is expressed in terms of gauge-invariant fields, and is displayed 
in these equations in a representation obtained in 
Ref.\cite{BCH3} by unitarily transforming the standard form
of that Hamiltonian.  But that does not change the fact that
${\hat H}_{\sf GI}$ is still the temporal-gauge Hamiltonian. 
In principle, ${\hat H}_{\sf GI}$ could be used to derive the equations
of motion of both, the gauge-invariant and the standard gauge-dependent temporal gauge fields, although the 
calculation leading to the latter would be clumsy and exceedingly tedious 
in this representation. In such calculations, we could not 
neglect the contributions made by commutators of operator-valued fields with 
$H_{\cal G}$, since the latter is an  
indispensable part of the temporal-gauge Hamiltonian in this particular representation. In reference to
Eqs. (\ref{eq:GHQCDN})-(\ref{eq:HamGauss}), ${H}_{\sf GI}$ and $H_{\cal G}$ --- 
the two constituent parts of ${\hat H}_{\sf GI}$  
--- can be given the following interpretation: ${H}_{\sf GI}$ can be recognized as a representation of the QCD 
Hamiltonian in the Coulomb gauge, similar to the Coulomb-gauge 
Hamiltonians obtained by methods different from 
ours.~\cite{schwingera,schwingerb,christlee,lee,creutz} 
These works differ among themselves and from ours in various ways 
--- in operator order, in the inclusion or omission of the 
longitudinal component of the chromoelectric field, in whether
the chromoelectric field and the glue color-charge density 
are hermitian --- but the Hamiltonians given by these authors are 
markedly similar to our ${H}_{\sf GI}$. Furthermore, 
$A^a_{{\sf GI}\,i}$ and ${\Pi}_{{\sf GI}\,j}^d$ respectively have the
same commutation rules with each other, and the components of 
${\Pi}_{{\sf GI}\,j}^d$ among themselves, as do the 
corresponding Coulomb-gauge fields in  Ref. \cite{schwingera}, {\em modulo}
operator ordering. \bs 

$H_{\cal G}$  has no role in the time evolution of physical 
systems, since it 
annihilates all states that implement Gauss's law  
(defined by ${\cal G}^a_{\sf GI}({\bf r})\,|{\Psi}\rangle=0$). When ${\cal G}_{\sf GI}^a$   
annihilates a state $|\Psi\rangle$, and thus identifies it as 
implementing Gauss's law,  it also annihilates
$\exp(-i{\hat H}_{\sf GI}t)|\Psi\rangle$. As we have shown, if $H_{\cal G}$ appears 
within a sequence of operators constructed from 
component parts of ${\hat H}_{\sf GI}$, such as might be 
found in an expansion of $\exp(-i{\hat H}_{\sf GI}t)$ or 
in a transition amplitude, and if this operator sequence 
acts on a state $|\Psi\rangle$ that satisfies ${{\cal G}}^a_{\sf GI}({\bf{x}})|\Psi\rangle=0$, then the 
resulting expression must vanish. 
${\cal G}_{\sf GI}^a$ can be commuted through the other operator to its right, 
producing further operators, each of which has 
a ${{\cal G}}^a_{\sf GI}$ on its right-hand side, until, finally 
all resulting terms have a ${{\cal G}}^a_{\sf GI}$ 
operator for some value of $a$ and some spatial argument, acting on 
$|\Psi\rangle$, with the result that the 
original expression ($i.\,e.$ the sequence of operators acting on $|\Psi\rangle$) vanishes.  
Finally, we note that ${\hat H}_{\sf GI}$ is not affected by the 
operator-ordering problem associated with the direct
quantization of QCD in the Coulomb gauge referred to in 
Ref.~\cite{christlee}.
The gauge-invariant field and negative chromoelectric field, 
$A^a_{{\sf GI}\,i}$ and $\Pi^b_{{\sf GI}\,j}$ respectively, 
obey commutation rules derived from their structure in 
terms of standard gauge-dependent temporal-gauge fields, whose commutation rules
are simple enough to make the commutation relations of $A^a_{{\sf GI}\,i}$ and $\Pi^b_{{\sf GI}\,j}$,
as well as their positions in the QCD Hamiltonian, well-defined; 
\bs

A remarkably similar state of affairs obtains in QED. When QED is formulated in the temporal gauge, and 
a unitary transformation is carried out which is the Abelian analog to the one that takes QCD in the temporal
gauge from the ${\cal C}$ to the ${\cal N}$ transformation as discussed in Ref.~\cite{CBH2}, the following 
result is obtained:\cite{khqedtemp,khelqed} The QED Hamiltonian in the temporal gauge, 
unitarily transformed by the Dirac 
transformation,\cite{diracgauge} can be described as
\bea
{\hat H}_{QED}={\int}d{\bf r}&&\!\!\!\!\!\!\!\!\left[{\textstyle \frac{1}{2}}{\Pi}_i({\bf r}){\Pi}_i({\bf r})+
{\textstyle \frac{1}{4}}F_{ij}({\bf r})F_{ij}({\bf r})+
{\psi^\dagger}({\bf r})\left(\beta m-i\alpha_{i}\partial_{i}\right)\psi({\bf r})\right]\nonumber \\
-&&\!\!\!\!\!\!\!\!{\int}d{\bf r}A^{(T)}_i({\bf r})j_i({\bf r})
+{\int}d{\bf r}d{\bf r}^{\prime} \frac{j_0({\bf r})
j_0({\bf r}^{\prime})}
{8{\pi}|{\bf r}-{\bf r}^{\prime}|}+H_g
\label{eq:Hqedtc}
\eea
where $A^{(T)}_i$ designates the transverse Abelian gauge field, and $H_g$ can be expressed as 
\be
H_g=-{\textstyle \frac{1}{2}}{\int}d{\bf r}\left({\partial}_i{\Pi}_i({\bf r})\frac{1}{\partial^2}j_0({\bf r})+
j_0({\bf r})\frac{1}{\partial^2}{\partial}_i{\Pi}_i({\bf r})\right)\,,
\label{eq:Hg}
\ee
so that all the operator-valued fields that appear in ${\hat H}_{QED}$ are gauge invariant, and so that
${\hat H}_{QED}$ also consists of two parts: the Hamiltonian for QED in the Coulomb gauge and 
$H_g$, which has no affect on the time evolution of states that implement Gauss's law (which,
in analogy to the non-Abelian ${\cal G}^a_{\sf GI}({\bf r})\,|{\Psi}\rangle=0$, takes the form 
${\partial}_i{\Pi}_i({\bf r})|\Phi\rangle=0$ after the Dirac transformation has 
been carried out).\footnote{In Refs.~\cite{khqedtemp,khelqed}, ${\partial}_i{\Pi}_i$
is separated into positive and negative frequency parts, so that the gauge-invariant charged states is 
manifestly normalizable. The unitary transformation in 
these references therefore differs somewhat from 
the one introduced by Dirac, but has essentially the same effect.}   
When the time derivative is defined as $i[{\hat H}_{QED}\,,]$, the equations of motion of temporal-gauge
QED result. But since $H_g$ can have no effect on the time evolution 
of state vectors that implement Gauss's law, 
the only part of ${\hat H}_{QED}$ that time-evolves such state vectors is the Coulomb-gauge Hamiltonian. 
This remarkable parallelism between QCD and QED  prevails because $V_{\cal C}$, which 
we constructed in Ref.\cite{CBH2}, is the 
non-Abelian generalization of the transformation which Dirac constructed for QED.

\section{Multiple solutions of the resolvent field equations and Gribov copies}
\label{sec:gribov}
The fact that we can identify the Coulomb-gauge field with the gauge-invariant 
fields we have constructed,\cite{CBH2} 
enables us to use previously reported results about the topology of the gauge-invariant 
gauge field~\cite{HCC}
to observe how the Gribov ambiguity manifests itself in the temporal-gauge formulation we have been 
discussing. \bs

The Gribov ambiguity --- the existence of multiple copies of fields that 
obey the Coulomb gauge condition --- complicates the quantization of QCD in the Coulomb gauge. 
This complication manifests itself as an inability to uniquely invert the Faddeev-Popov 
operator $D_i\partial_i$. Gauss's law in the Coulomb gauge
can be written in the form $D_i\partial_iA_0^a=-j_0^a$, in which the Faddeev-Popov operator
takes account of the fact that 
the ``total'' color charge-density (including quark and 
glue color contributions) is $J_0^a=g\epsilon^{abc}A_i^b\Pi_i^c+j_0^a$. 
It is necessary to invert $D_i\partial_iA_0^a=-j_0^a$ in order 
to replace $A_0^a$ in the Coulomb-gauge Hamiltonian with its equivalent 
in terms of unconstrained field variables. 
This inversion produces not only operator-ordering problems; it also 
leads to the Gribov problem. Inverting $D_i\partial_iA_0^a=-j_0^a$ formally produces the equation 
$A_0^a=-{(D_i\partial_i)}^{-1}j_0^a$, the non-Abelian analog of $A_0=-{\partial}^{-2}j_0$ in QED. 
However, whereas ${\partial}^{2}$ is trivially
invertible for suitably chosen boundary conditions, and manifestly commutes with the Abelian
gauge field as well as with its conjugate momentum, 
the differential operator $D_i\partial_i$ is not similarly invertible and does not commute with the 
non-Abelian gauge field or with the chromoelectric field. 
$A_0^a$ therefore cannot be uniquely replaced by an expression that depends on unconstrained field 
variables only.\bs 

Various authors have responded to this 
complication in different ways. Some have concluded that the difficulties associated with the Coulomb 
gauge make it preferable to use an axial gauge, such as the temporal gauge, 
in which, they state, the Gribov problem does not exist.\cite{wein} 
Other authors have taken the point of view that, since it is likely that  Gribov copies
of the original gauge field belong to different topological sectors, one can ignore the Gribov
copies by limiting oneself to the perturbative regime.~\cite{ramond} It has also been questioned 
whether the Gribov ambiguity is truly absent when axial gauges are imposed, or whether it
always affects non-Abelian gauge theories, but is more obscure and harder to detect in the case
of axial gauges. Singer has proven that when the path integral of a non-Abelian gauge theory 
is defined over a Euclidean 4-dimensional sphere, ambiguities appear
regardless of the gauge condition; but, as Singer has noted, 
in the absence of such boundary conditions, there are gauges ---
in particular axial gauges --- to which his proof does not apply.\cite{Singer}
Others have argued that given these circumstances, it is still
uncertain whether Gribov ambiguities are endemic to non-Abelian gauge theories or whether they only 
mark the Coulomb gauge as an unfortunate choice of gauge.\cite{ramond}  \bs

In previous work\cite{CBH2,BCH3,CHQC,HCC} and in previous sections of this paper, we have given series 
representations of the resolvent field as well as of the inverse of the Faddeev-Popov operator
and of a number of Dirac commutators. We have also shown that even when 
functionals of gauge fields are given a series representation, that does not 
necessarily eliminate the possibility of representing them nonperturbatively. 
These functionals can be related by integral equations, and nonperturbative solutions of these
equations can be obtained. In this way, the origin of multiple solutions of gauge fields 
can be better understood\bs

 In Ref.\cite{HCC}, we made an {\em ansatz}, representing the gauge field in the temporal gauge and the 
resolvent field  as
functions of spatial variables that are second-rank tensors 
in the combined spatial and SU(2) indices; except in so far as 
the forms of $\overline{{\cal{A}}_{i}^{\gamma}}({\bf{r}})$ and 
$A_i^{\gamma}({\bf r})$ reflect this second-rank 
tensor structure, they are represented as 
isotropic functions of position. In this way, we can represent the 
longitudinal part of the gauge field in the temporal gauge as  
 \begin{equation}
A_i^{\gamma}\,^L({\bf r})=\frac{1}{g}\left[{\delta}_{{i}\,{\gamma}}\,\frac{{\cal N}(r)}{r}+
\frac{r_i\,r_{\gamma}}{r}\,\left(\frac{{\cal N}(r)}{r}\right)^{\prime}\,\right]
\label{eq:longa}
\end{equation}
and the transverse part as
\begin{equation}
A_i^{\gamma}\,^T({\bf r})={\delta}_{{i}\,{\gamma}}\,{\cal T}_{A}(r)+
\frac{r_i\,r_{\gamma}}{r^2}\,{\cal T}_{B}(r)+{\epsilon}_{i{\gamma}n}\frac{r_n}{r}\,{\cal T}_{C}(r)
\label{eq:transa}
\end{equation}
where ${\cal N}(r)$, ${\cal T}_{A}(r)$, ${\cal T}_{B}(r)$ and ${\cal T}_{C}(r)$ are isotropic
functions of $r$, the prime denotes differentiation with respect to $r$, and 
the transversality of $A_i^{\gamma}\,^T({\bf r})$ requires that
\begin{equation}
\frac{d(r^2{\cal T}_B)}{dr}+r^2\frac{d\,{\cal T}_A}{dr}=0\,.
\label{eq:trans}
\end{equation}
An entirely analogous representation of the resolvent field enables us to relate it
to the gauge field through the nonlinear integral equation described in Section \ref{sec:intro}. 
As a result of this analysis, we have
been able to show that it is possible to
represent the resolvent field as~\cite{HCC}
\begin{equation}
\overline{{\cal{A}}_{i}^{\gamma}}({\bf{r}})=\left({\delta}_{i\,{\gamma}}-\frac{r_i\,r_{\gamma}}{r^2}\right)
\left(\frac{\overline{\cal{N}}}{gr}+{\varphi}_A\right)+{\epsilon}_{i{\gamma}n}\frac{r_n}{r}\,{\varphi}_C
\label{eq:aform}
\end{equation} 
where 
\be
\varphi_A=\frac{1}{gr}\left[{\cal{N}}{\cos}(\overline{\cal{N}}+{\cal{N}})-{\sin}(\overline{\cal{N}}+
{\cal{N}})\right]
+{\cal T}_A\left[{\cos}(\overline{\cal{N}}+{\cal{N}})-1\right]-
{\cal T}_C\,{\sin}(\overline{\cal{N}}+{\cal{N}})
\label{eq:phia}
\ee 
and
\begin{equation}
\varphi_C=\frac{1}{gr}\left[{\cal{N}}{\sin}(\overline{\cal{N}}+{\cal{N}})+
{\cos}(\overline{\cal{N}}+{\cal{N}})-1\right]
+{\cal T}_C\left[{\cos}(\overline{\cal{N}}+{\cal{N}})-1\right]+
{\cal T}_A\,{\sin}(\overline{\cal{N}}+{\cal{N}}).
\label{eq:phic}
\end{equation}
Similarly, the gauge-invariant gauge field can be expressed as 
a functional of $\overline{\cal{N}}$ and of ${\cal{N}}$, ${\cal T}_A$, ${\cal T}_B$ and ${\cal T}_C$
as shown by  
\begin{eqnarray}
&&A_{{\sf GI}\,i}^{\gamma}({\bf{r}})=\frac{1}{gr}\left\{\epsilon_{i\,\gamma\,n}\frac{r_n}{r}
\left[{\cos}(\overline{\cal{N}}+{\cal{N}})-1+{\cal{N}}\sin(\overline{\cal{N}}+{\cal{N}})\right]
+\left({\delta}_{i\,\gamma}-\frac{r_ir_{\gamma}}{r^2}\right)\times\right.\nonumber\\
&&\left.\times\left[{\cal{N}}{\cos}(\overline{\cal{N}}+{\cal{N}})-{\sin}(\overline{\cal{N}}+{\cal{N}})\right]
-\frac{r_ir_{\gamma}}{r}\frac{d\overline{\cal{N}}}{dr}\right\}\nonumber \\
&&+{\cal T}_A\left\{
\left({\delta}_{i\,\gamma}-\frac{r_ir_{\gamma}}{r^2}\right)
{\cos}(\overline{\cal{N}}+{\cal{N}})
+\epsilon_{i\,\gamma\,n}\frac{r_n}{r}{\sin}(\overline{\cal{N}}+{\cal{N}})\right\}+\frac{r_ir_{\gamma}}{r^2}
\left({\cal T}_A+{\cal T}_B\right)\nonumber \\
&&+{\cal T}_C\left\{\epsilon_{i\,\gamma\,n}\frac{r_n}{r}{\cos}(\overline{\cal{N}}+{\cal{N}})
-\left({\delta}_{i\,\gamma}-\frac{r_ir_{\gamma}}{r^2}\right)
{\sin}(\overline{\cal{N}}+{\cal{N}})\right\}.
\label{eq:AGIsub}
\end{eqnarray}
With these representations, 
the nonlinear integral equation that relates the resolvent field to the gauge field (in the temporal-gauge)
is transformed to the nonlinear differential equation  
\begin{eqnarray}
\frac{d^2\,\overline{\cal{N}}}{du^2}&+&\frac{d\,\overline{\cal{N}}}{du}+
2\left[{\cal{N}}{\cos}(\overline{\cal{N}}+{\cal{N}})-
{\sin}(\overline{\cal{N}}+{\cal{N}})\right]\nonumber \\
&+&2gr_0\exp(u)\left\{{\cal T}_A\left[{\cos}(\overline{\cal{N}}+{\cal{N}})-1\right]-{\cal T}_C\,
{\sin}(\overline{\cal{N}}+{\cal{N}})\right\}=0
\label{eq:nueq}
\end{eqnarray}
where $u=\ln(r/r_0)$ and $r_0$ is an arbitrary constant. $\overline{\cal{N}}$ --- the dependent variable in 
Eq. (\ref{eq:nueq}) --- can be directly related to the resolvent
field $\overline{{\cal{A}}_{i}^{\gamma}}$ by 
\be
\overline{\cal{N}}=g\frac{r_{\alpha}}{r}
\frac{\partial_{j}}{\partial^{2}}\overline{{\cal A}_{j}^{\alpha}}({\bf r}).
\label{eq:nbar}
\ee
Similarly, ${\cal{N}}$ can be related to the gauge field in the temporal (Weyl) gauge as shown by 
\be
{\cal N}=g\frac{r_\alpha}{r}
\frac{\partial_{j}}{\partial^{2}}A_{j}^{\alpha}({\bf r}).
\label{eq:nweyl}
\ee
 ${\cal T}_A$, ${\cal T}_B$ and ${\cal T}_C$ are determined by Eq. (\ref{eq:trans}) and by 
the transverse part of the gauge field as shown by
\be
\frac{r_ir_{\gamma}}{r^2}A_i^{\gamma}\,^T={\cal T}_A+{\cal T}_B
\;\;\;\mbox{and by}\;\;\;\frac{1}{2}{\epsilon}^{i{\gamma}j}\frac{r_j}{r}A_i^{\gamma}\,^T={\cal T}_C\,.
\label{eq:tatb}
\ee
In solving Eq. (\ref{eq:nueq}), boundary conditions are imposed on 
$\overline{\cal{N}}$ and on the source terms 
${\cal N}$, ${\cal T}_{A}$, ${\cal T}_{B}$ and ${\cal T}_{C}$. 
${\cal T}_{A}$, ${\cal T}_{B}$ and ${\cal T}_{C}$ 
are required to vanish as $u\ra\pm\infty$; ${\cal N}$ is required to vanish as $u{\ra}-\infty$,
and to either vanish or approach specified limits as $u{\ra}\infty$. 
$\overline{\cal{N}}$ is required to be bounded 
in the entire interval $-\infty{\leq}u{\leq}\infty$. For convenience, 
we normalize $\overline{\cal{N}}$ so that
$\overline{\cal{N}}{\ra}0$ when $u{\ra}-\infty$. 
\bs

One of the questions addressed in Ref.\cite{HCC} is the following: given a specified set of values
${\cal N}(r)$, ${\cal T}_{A}(r)$, ${\cal T}_{B}(r)$ and ${\cal T}_{C}(r)$, what behavior is possible for  
$\overline{\cal{N}}$, if $\overline{\cal{N}}$ is required to be bounded 
in the entire interval $0{\leq}r\!<\infty$
and to obey Eq. (\ref{eq:nueq})? In posing this question, 
$\overline{\cal{N}}$ is seen to be the unknown function, and   
${\cal N}(r)$, ${\cal T}_{A}(r)$, ${\cal T}_{B}(r)$ and 
${\cal T}_{C}(r)$ are source terms that drive the solution, 
subject to the required boundary conditions. In Ref.~\cite{HCC}, we presented 
 numerical integrations of Eq.~(\ref{eq:nueq}) which demonstrated that a number of different
solutions of this equation --- and, hence, a number of different $A_{{\sf GI}\,i}^{\gamma}$ ---
can be based on the same gauge-dependent gauge field characterized by a single set of values of 
${\cal N}$, ${\cal T}_{A}$, ${\cal T}_{B}$ and ${\cal T}_{C}$. A related 
question --- what form for $A_{{\sf GI}\,i}^{\gamma}({\bf{r}})$
corresponds to the different possible functional forms of $\overline{\cal{N}}$ --- can also be posed.  
When the gauge field $A^{\gamma}_i=0$, the resolvent 
field ${\bar {\cal A}}^{\gamma}_i$ and the gauge-invariant field
$A^{\gamma}_{{\sf GI}\,i}$ need not vanish. In that case,
Eq. (\ref{eq:nueq}) reduces to the autonomous\footnote{Eq. (\ref{eq:nueqzero}) is autonomous because the  
variable $u$ does not appear explicitly in this equation; it appears implicitly only as the argument of 
$\overline{\cal{N}}$ and of its derivatives.}
 ``Gribov equation''~\cite{gribov,gribovb}
\begin{equation}
\frac{d^2\,\overline{\cal{N}}}{du^2}+\frac{d\,\overline{\cal{N}}}{du}-
2{\sin}(\overline{\cal{N}})=0
\label{eq:nueqzero}
\end{equation}
 which is also the equation for a damped pendulum with $\overline{\cal{N}}$ representing the angle with
respect to the pendulum's position of unstable equilibrium, and $u$ representing the time.\footnote{the 
function $\alpha$ in Ref.\cite{gribovb} is related to $\overline{\cal{N}}$ by $\overline{\cal{N}}=2\alpha$.}  
In the application of this equation to the Gribov problem, 
$\overline{\cal{N}}$ must remain bounded not only in the interval $0{\leq}u<\infty$, but also
in the larger interval $-{\infty}<u<\infty$ to include 
the entire configuration space $0{\leq}r<\infty$. \bs

The restriction that $\overline{\cal{N}}$ must be bounded in the 
entire interval $(-\infty{\leq}u{\leq}\infty)$
severely limits the allowed solutions of Eq. (\ref{eq:nueqzero}). For solutions of Eq. (\ref{eq:nueqzero})
that are bounded in the entire interval $(-\infty{\leq}u{\leq}\infty)$, there is a single, 
unique, phase plot. One branch extends from the unstable saddle point (corresponding to 
$u{\rightarrow}-\infty)$, at which $\overline{\cal{N}}=0$, to a stable point (corresponding to 
$u{\rightarrow}\infty)$, at which $\overline{\cal{N}}=\pi$. The other branch extends from 
the same saddle point (at which $u$ corresponds to $-\infty$) to a stable point at $\overline{\cal{N}}=-\pi$,
also corresponding to $u{\rightarrow}\infty$. The two branches are identical, except 
that for each point on one branch
the values of $\overline{\cal{N}}$ and $d\overline{\cal{N}}/du$ correspond to values 
 $-\overline{\cal{N}}$ and $-d\overline{\cal{N}}/du$ on the other.
Numerical solutions are obtained by setting $\overline{\cal{N}}$
and $d\overline{\cal{N}}/du$ equal to the same small value at some large negative 
value of $u$, in order to discriminate against solutions that become unbounded as $u$ approaches the 
saddle point as $u{\rightarrow}-\infty$, as discussed in Ref.\cite{HCC}. We can choose different
large negative values of $u$ at which to set $\overline{\cal{N}}=d\overline{\cal{N}}/du=0$; if,
in one case, we choose $u_a$ and in another $u_b$, we obtain solutions
$\overline{\cal{N}}_a(u)$ and $\overline{\cal{N}}_b(u)$ respectively, where  
$\overline{\cal{N}}_b(u)=\overline{\cal{N}}_a(u+{\sf U})$ and ${\sf U}=u_b-u_a$. 
Similarly, changing the magnitude of the small value 
of $\overline{\cal{N}}(u_a)= d\overline{\cal{N}}(u_a)/du=\epsilon$ 
to $\overline{\cal{N}}(u_a)= d\overline{\cal{N}}(u_a)/du={\epsilon}^{\,\prime}$ 
also has the effect of shifting the functional form of $\overline{\cal{N}}(u)$ from $u$ to  
$u+u_0$ for a particular value of $u_0$. Replacing the initial condition 
$\overline{\cal{N}}(u_a)=d\overline{\cal{N}}(u_a)/du=\epsilon$, for a small $\epsilon$, with
$\overline{\cal{N}}(u_a)=d\overline{\cal{N}}(u_a)/du=-\epsilon$, has no other effect than changing the 
signs of $\overline{\cal{N}}(u)$ and $d\overline{\cal{N}}/du$. 
Figure 1 illustrates some of these relationships.\bs

To study the variation in the form of $\overline{\cal{N}}$ that is allowed when $\overline{\cal{N}}$ 
is represented in configuration space in the form $\overline{\cal{N}}(u(r))=\overline{\cal{N}}(\ln(r/r_0))$,
we observe the following: For the shift $\overline{\cal{N}}_b(u)=\overline{\cal{N}}_a(u+{\sf U})$ with
${\sf U}$ represented as ${\sf U}=\ln(r_0/R_0)$ for an appropriate $R_0$,
\be
\overline{\cal{N}}_b\left(\ln(r/r_0)\right)=\overline{\cal{N}}_a\left(\ln(r/r_0)+\ln(r_0/R_0)\right)
=\overline{\cal{N}}_a\left(\ln(r/R_0)\right).
\label{eq:NaNb}
\ee  
It is clear that when $\overline{\cal{N}}$ is required to be bounded in the entire interval
$(0{\leq}r{\leq}\infty)$, a change in the constant $r_0$ and an overall change of sign are the only changes 
allowed in $\overline{\cal{N}}\left(\ln(r/r_0)\right)$. In particular, in {\em every} such case, 
$\overline{\cal{N}}(r=0)=0$ and $\overline{\cal{N}}(r{\rightarrow}{\infty})=\pm\pi$. When the 
restrictions on ${\cal{N}}$, ${\cal T}_A$, ${\cal T}_B$ and ${\cal T}_C$ that led to 
Eq. (\ref{eq:nueqzero}) are applied to Eq. (\ref{eq:AGIsub}), we find that 
\be
\left[A_{{\sf GI}\,i}^{\gamma}({\bf{r}})\right]_{(0)}=\frac{-2}{gr}\epsilon_{i\,\gamma\,n}\frac{r_n}{r},
\label{eq:AGIsubzero}
\ee
where the subscript $(0)$ designates the solution that corresponds to the ``pure gauge'' case in which 
the gauge-dependent gauge field $A_i^{\gamma}$ vanishes. $[A_{{\sf GI}\,i}^{\gamma}({\bf{r}})]_{(0)}$
can easily be recognized as a ``hedgehog'' solution.\bs

Some authors have suggested that Eq. (\ref{eq:nueqzero}) --- the Gribov equation ---
has a variety of solutions for which 
$\overline{\cal{N}}(r{\rightarrow}{\infty})$ can be any integer multiple of $\pi$,~\cite{kaku} and 
that different integer multiples correspond to different topological sectors connected by 
large gauge transformations. This suggestion, which is motivated by the model of a damped 
pendulum, neglects the fact that in the Gribov equation $u$ must be bounded in the entire interval
$(-\infty{\leq}u{\leq}\infty)$, and that the only solutions of the damped pendulum problem that 
can remain bounded in this entire interval as the time $u$ is extrapolated backwards to $-\infty$, 
are those for which the pendulum initially is at rest in 
its unstable equilibrium position. And, with this initial position, the damped pendulum is unable to 
execute multiple turns before coming to rest at equilibrium in a stable configuration.\bs

Gribov explicitly noted that the necessity of requiring the solutions to 
Eq. (\ref{eq:nueqzero}) to be bounded
in the entire interval $-{\infty}<u<\infty$ limits the asymptotic values of 
$\overline{\cal{N}}(u{\rightarrow}{\infty})$ to $\pm\pi$.\cite{gribovb} He therefore also 
considered the transverse gauge field that results when an arbitrarily chosen transverse gauge field is 
gauge-transformed, so that the new field is not pure gauge, but is given by 
\be
{\sf A}^{\prime}_i=U{\sf A}_iU^{-1}+iU\partial_iU^{-1}          
\label{eq:grib_f}
\ee
$$\mbox{where}\;\;U=\exp\left(-i{\phi}(r)\frac{{\vec r}\cdot
{\vec \tau}}{2r}\right)\;\;\mbox{and}\;\;{\sf A}_i={\sf A}^c_i\frac{{\tau}^c}{2}
\;\;\mbox{with}\;\;{\sf A}^c_i={\epsilon}^{ijc}\frac{r_j}{r^2}f(r).$$ 
${\sf A}^{\prime}_i$ and ${\sf A}_i$ both are transverse and 
therefore belong to the Coulomb gauge. The transversality
of ${\sf A}^{\prime}_i$ leads to the equation 
\be
\frac{d^2\,\phi}{du^2}+\frac{d\,\phi}{du}-2{\sin}(\phi)\left(1-f(u)\right)=0
\label{eq:gribphi}
\ee
which is not autonomous, and does have the multiple solutions that 
correspond to Gribov copies belonging to different topological sectors.\bs

In our work, Eq. (\ref{eq:nueq}) --- of which Eq. (\ref{eq:nueqzero}) is a special case --- also has 
inhomogeneous source terms that prevent it from being autonomous. But, unlike Eq. (\ref{eq:gribphi}),
Eq. (\ref{eq:nueq}) is not obtained by gauge-transforming from one Coulomb-gauge field to another. It 
is a consequence of the transformation from ``standard'' gauge-dependent fields in the temporal gauge 
to the corresponding gauge-invariant fields. Our Eq. (\ref{eq:nueqzero}) describes this transformation
for the case that the initial gauge-dependent temporal-gauge field is set $=0$, which clearly corresponds
to the ``pure gauge'' case --- Gribov's Eq. (\ref{eq:gribphi}) with $f(u)=0$.  But, more generally, 
the relation between our Eqs. (\ref{eq:nueq}) and (\ref{eq:nueqzero}) is different from 
the relation of Gribov's Eq.~(\ref{eq:gribphi}) to that same equation with $f(u)=0$.
Our Eq. (\ref{eq:nueq}) would not
reduce to an equation of the form of Eq. (\ref{eq:gribphi}) even if the initial gauge-dependent field 
were chosen to be purely transverse. Eq. (\ref{eq:nueq}) shows that, in that case,
 there would be the additional ``source''-term $2gr_0\exp(u){\cal T}_A({\cos}\,\overline{\cal{N}}-1)$ 
(where $\overline{\cal{N}}$
corresponds to $\phi$ in Eq. (\ref{eq:gribphi})). Nevertheless, in spite of these differences,
the fact that we are able to identify the gauge-invariant 
gauge fields we constructed with the Coulomb-gauge fields enables us to interpret  the multiple solutions of 
Eq. (\ref{eq:nueq}), which we demonstrated in Ref.~\cite{HCC},
as Gribov copies.  These multiple solutions represent Gribov copies of the gauge-invariant gauge fields in the 
temporal gauge, which manifest themselves even when there are no such copies, 
and no ambiguity, in the case of the ``standard'' gauge-dependent temporal-gauge fields.

\section{Discussion}
\label{sec:disc}
In this work, we have established a relationship between the gauge-invariant 
temporal-gauge and the Coulomb-gauge formulations of two-color QCD: The temporal-gauge QCD 
Hamiltonian, when represented entirely in terms of
gauge-invariant operator-valued fields, is not identical to the Coulomb-gauge Hamiltonian 
represented by $H_{\sf GI}$  given in Eq. (\ref{eq:HQCDN}). But the two are physically 
equivalent --- $i.\,e.$ they lead to identical values of observable quantities within 
the space of states in which Gauss's law has been implemented. Moreover, we have shown that the 
gauge-invariant temporal-gauge fields obey
commutation rules that are the same as those for Coulomb-gauge fields, 
{\em modulo} operator-ordering ambiguities,
when the longitudinal components of the Coulomb-gauge chromoelectric field are retained,  
as they are in our work and in Schwinger's treatment of the Coulomb gauge.\cite{schwingera}\bs
 
We have also shown, through reference to specific numerical calculations,\cite{HCC} 
that there are Gribov copies
of {\em gauge-invariant~} fields in the temporal-gauge formulation of QCD, 
even though there are no Gribov copies of
the {\em gauge-dependent~} temporal-gauge fields. We have demonstrated 
that the gauge-invariant fields in the temporal gauge obey a nonlinear integral equation, 
which --- subject to an {\em ansatz} --- can be transformed to a 
nonlinear differential equation  that has multiple
solutions corresponding to a single gauge-dependent field. 
These solutions must be bounded in the entire 
configuration space. Thus, the nonlinear differential equation that embodies the 
imposition of Gauss's law and the implementation of gauge invariance in the temporal gauge,
and the requirement of boundedness, lead to the multiple solutions that can be identified
as Gribov copies. This is consistent with our demonstration of a close resemblance between 
the gauge-invariant formulation of the temporal gauge and the Coulomb-gauge formulation of QCD.
It is also consistent with the fact that the nonlinear differential equation that relates the
gauge-invariant and the gauge-dependent temporal-gauge fields --- Eq. (\ref{eq:nueq}) ---  
has a form very similar to the one 
that Gribov used to demonstrate the non-uniqueness of Coulomb-gauge fields 
--- Eq. (\ref{eq:gribphi}).\bs

We are therefore led to conclude that the reason why
the Gribov ambiguity does not arise when QCD 
is quantized in the temporal gauge --- and most authors who discuss the quantization of 
QCD in the temporal gauge either do not mention the Gribov ambiguity or state that 
there is no Gribov ambiguity in the temporal gauge~\cite{wein} --- is  that,
in sharp contrast to the Coulomb gauge,  quantization in the temporal  
gauge proceeds to completion without requiring the imposition of Gauss's law. 
The Gribov copies arise in 
the temporal-gauge formulation only after the theory has been quantized, and 
Gauss's law then is implemented and gauge-invariant fields are constructed.  
Our results are consistent with the conclusion that the Gribov 
ambiguity is a fundamental attribute of non-Abelian gauge theories. Gribov copies do not 
arise when a non-Abelian gauge theory is originally quantized in the 
temporal or other axial gauges, because, then, 
Gauss's law remains unimplemented. The Gribov ambiguity does
manifest itself in these gauges, but only with the introduction of the 
gauge-invariant fields and the imposition of Gauss's law.

\section{Acknowledgements}
The author thanks Profs. Hai-cang Ren and Y. S. Choi for helpful conversations. 
This research was supported by the Department of Energy under Grant
No.~DE-FG02-92ER40716.00.\bs \bs 
 
\n
{\Large {\bf Appendix}}\bs

\n
In this Appendix, we will evaluate the quantities ${\cal P}_{ui}$ and ${\cal R}_{vb}$ defined
in Eq. (\ref{eq:demapp}), and show that they have the required properties. \bs

For the case of ${\cal P}_{ui}$, we evaluate  $V_{\cal{C}}\partial_iV_{\cal{C}}^{-1}$,
where $V_{\cal{C}}$ is most conveniently represented as shown in Eq. (\ref{eq:vcz}),
so that 
\be
\chi_i=V_{\cal{C}}\partial_iV_{\cal{C}}^{-1}=\exp\left(-ig{\cal Z}^\alpha({\bf{r}})
{\textstyle\frac{\tau^\alpha}{2}}\right)\partial_i\exp\left(ig{\cal Z}^\alpha({\bf{r}})
{\textstyle\frac{\tau^\alpha}{2}}\right).
\label{eq:psione}
\ee
 For $\Phi^{\alpha}=g{\cal Z}^{\alpha}$ and $\Phi=\sqrt{\Phi^{\alpha}\Phi^{\alpha}}$,~ we obtain 
\be
\chi_i=\left\{\cos\left(\frac{{\Phi}}{2}\right)-i\frac{\tau^a{\Phi}^a}
{{\Phi}}\sin\left(\frac{{\Phi}}{2}\right)\right\}\partial_i
\left\{\cos\left(\frac{{\Phi}}{2}\right)+i\frac{\tau^a{\Phi}^a}
{{\Phi}}\sin\left(\frac{{\Phi}}{2}\right)\right\}.
\label{eq:psitwo}
\ee
 Eq. (\ref{eq:demapp}) then determines that
\be
{\cal P}_{ai}=\partial_i{\Phi}^a+\left(\frac{\Phi^a\partial_i\Phi}{\Phi}-\partial_i{\Phi}^a\right)
\left(1-\frac{\sin{\Phi}}{\Phi}\right)+\epsilon^{abc}\Phi^b\partial_i{\Phi}^c
\left(\frac{1-\cos{\Phi}}{\Phi^2}\right).
\label{eq:psithree}
\ee
Similarly, Eq. (\ref{eq:demapp}) also determines that ${\cal R}_{ba}$ is given by
\be
{\cal R}_{ba}=\delta_{ab}\cos{\Phi}+{\Phi}^a{\Phi}^b\left(\frac{1-\cos{\Phi}}{{\Phi}^2}\right)
+\epsilon^{abc}{\Phi}^c\left(\frac{\sin{\Phi}}{{\Phi}}\right).
\label{eq:Tab}
\ee
Inspection of Eq. (\ref{eq:Tab}) demonstrates that simultaneous substitution of $-{\vec {\Phi}}$ 
for ${\vec {\Phi}}$ and exchange of the subscripts $a$ and $b$ in ${\cal R}_{ba}$ leaves the expression
on the right-hand side of this equation unchanged, thus proving Eq. (\ref{eq:APj0}). \bs \bs

\newpage
\input{epsf.tex}

\begin{figure}[ht]
\begin{center}
\leavevmode
\epsfxsize=6in
\includegraphics[scale=1.00]{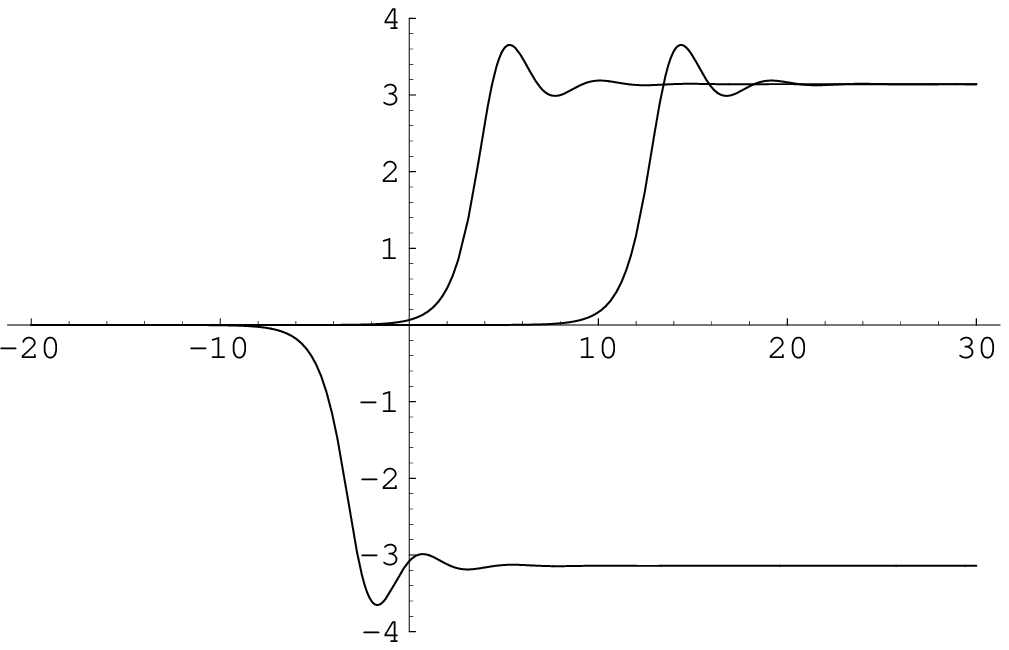}
\end{center}
\caption{\label{fig1}}
\end{figure}

\begin{center}{Figure Caption}\end{center}
\bs

\n
Three solutions of Eq. (\ref{eq:nueqzero}) with $\overline{\cal{N}}$ bounded in the interval 
$(-\infty{\leq}u{\leq}\infty)$. Of the two solutions with $\overline{\cal{N}}{\geq}0$, 
one is obtained with 
$\overline{\cal{N}}=d\overline{\cal{N}}/du=10^{-6}$ at $u=-11$ and the other with
$\overline{\cal{N}}=d\overline{\cal{N}}/du=10^{-10}$ at $u=-11$. The
plot for the latter solution is identical to the plot of the former 
shifted to the right on the $u$ axis. Both solutions approach $\overline{\cal{N}}=\pi$
as $u{\rightarrow}\infty$. The solution with $\overline{\cal{N}}{\leq}0$ 
is obtained with $\overline{\cal{N}}=d\overline{\cal{N}}/du=-10^{-6}$
at $u=-18$.  In this solution, $\overline{\cal{N}}{\rightarrow}-\pi$ as $u{\rightarrow}\infty$.
The three solutions are precisely identical {\em modulo} shifts along the $u$ axis 
and reflection in it. 
\end{document}